\documentclass[12pt,a4paper]{article}
\usepackage[utf8]{inputenc}
\usepackage[T1]{fontenc}
\usepackage[english]{babel}
\usepackage{geometry}
\usepackage{algorithm}
\usepackage{algpseudocode}
\usepackage{hyperref}
\usepackage{graphicx}
\usepackage{booktabs}
\usepackage{longtable}
\usepackage{array}
\usepackage{enumitem}
\usepackage{listings}
\usepackage{xcolor}
\usepackage{amsmath}
\usepackage{caption}
\usepackage{float}
\usepackage{setspace}
\title{Web Price Extraction: State of the Art and an Adaptive Browserless Implementation}
\author{
  \small Department of Computer Science and Systems Engineering, University of Zaragoza,\\
  \small Zaragoza, Spain \\
  \Large Evgeniia Kositsyna \\
  \Large Jorge Lloret-Gazo
}
\date{}

\begin{document}

\maketitle

\begin{abstract}
Price extraction from websites is a key task for market monitoring, price comparison, and business analytics in e-commerce. Existing approaches can be broadly divided into four groups, and understanding their trade-offs in accuracy and scalability is essential for selecting suitable extraction strategies. Classical methods rely on manually written wrappers and rule induction from labeled pages, offering high accuracy but adapting poorly to structural changes and requiring considerable maintenance effort. Browser-based methods, using tools such as Selenium and Puppeteer, handle dynamic JavaScript content but consume large computational resources and scale poorly. Browserless approaches retrieve HTML directly via HTTP requests, offering significant gains in speed and cost, but rely on rules calibrated for specific sites. Methods based on machine learning and large language models offer adaptability but require training data and substantial computation.

Our main contribution is an adaptive browserless price extraction system that improves robustness to structural differences between websites. We implemented a baseline architecture combining HTML page fragmentation with syntactic, semantic, and frequency rules, and extended it in two ways: a Bayesian approach that dynamically updates rule weights, and a genetic algorithm that optimizes the system's global parameters. This hybrid scheme increased precision from 77.2\% to 87.3\% and reduced average per-page processing time by approximately 14\% relative to the baseline, confirming it as a competitive alternative to manually tuned browserless solutions and to more resource-intensive browser- or LLM-based methods, offering high extraction accuracy at low computational cost.
\end{abstract}

\section{Introduction}
Web data extraction is a method of automatically obtaining structured information from unstructured or semi-structured web sources. The continued growth of online content has made the automated transformation of web pages into structured data a relevant problem for business, scientific and industrial applications.

The global competitor price monitoring market was estimated at USD~1.2 billion in 2024 and is projected to reach USD~2.5 billion by 2033~\cite{tendem2026}, demonstrating the increasing adoption of automated competitor monitoring and price scraping technologies in e-commerce. Companies use price scraping to monitor competitors' pricing strategies, detect market trends, and automatically adjust their own prices in real time. A practical example of the importance of price monitoring is provided by the study of Assad et al.~\cite{assad2024algorithmic}, conducted on pricing data from all German gas stations (over 16,000 stations), some of which switched to algorithmic pricing. Since the exact date of software adoption is unknown, the authors identify the moment of transition through a test for structural breaks across three measures of pricing behaviour simultaneously: the frequency of price changes, the average size of price changes, and the speed of reaction to a competitor's price change. This revealed that, in non-monopoly markets, where a station faces competitors, the adoption of the algorithm increases margins by 9\%. In duopoly markets, margins do not change if only one of the two parties adopts the algorithm, but increase by 28\% if both switch to algorithmic pricing. Thus, algorithms competing with one another in real time, based on collected competitor price data, effectively converge to higher—that is, less competitive—prices, which has a significant impact on companies' economics. Such systems help online retailers remain competitive, optimise profit margins, and react quickly to changes in customer demand and competitor activity. Therefore web scraping techniques are currently applied in areas such as competitive business intelligence and e-commerce systems.

The objective of this work is to analyse and compare different web scraping methods used for price extraction on the web, evaluating their characteristics, advantages, and limitations. The increasing interest of companies and organisations in monitoring information about competitors' products and automating large-scale data collection has made web scraping an essential tool in many domains. However, the extraction of reliable information from websites remains a challenging problem due to the constant evolution of HTML structures, the widespread use of JavaScript-generated content, and the growing adoption of anti-scraping mechanisms.

Several approaches and tools have been developed to address web data extraction problems. Traditional methods, commonly known as wrappers, consist of manually designed extraction programs tailored to a specific website or domain. To use these methods, it is necessary to analyse the HTML structure beforehand and they are usually precise. Nevertheless, they present important limitations, such as poor adaptability to structural modifications, high maintenance costs, and significant manual effort when websites change their layouts.

Modern approaches have attempted to overcome these limitations and can generally be divided into three main categories. The first category includes browser-based methods that simulate user navigation through automated browsers such as Selenium or Puppeteer. These approaches are capable of interacting with dynamic content generated by JavaScript and complex user interfaces. However, they often require considerable computational resources, generate high network traffic, and may suffer from scalability issues.

The second category groups browserless methods, which directly access HTML content without rendering a complete browser environment. Tools such as Requests, BeautifulSoup, and Scrapy belong to this group. Browserless techniques reduce resource consumption, improve execution speed, and simplify large-scale extraction tasks~\cite{lloret2020}. However, they may face difficulties when dealing with highly dynamic websites or client-side rendering mechanisms.

Finally, methods based on artificial intelligence and machine learning have emerged as a promising alternative. These techniques aim to identify patterns automatically and adapt to changing webpage structures through machine learning models and natural language processing techniques~\cite{ferrara2010}. These approaches could lessen the need for manual configuration and enhance resilience to structural changes. However, they also introduce new challenges related to training complexity, dataset availability, computational cost, and model generalisation.

This work focuses particularly on the comparison between browserless solutions and machine learning approaches for web price extraction. The main idea behind this work is to evaluate whether lightweight browserless architectures can provide competitive performance compared to more adaptive machine learning techniques while maintaining lower computational requirements and implementation complexity.

The main contributions of this work are:

\begin{itemize}
  \item a comparative analysis of classical, browser-based, browserless, and AI-driven web scraping methods;
  \item the implementation in Python of a browserless price extraction system based on syntactic and semantic rules;
  \item the extension of the baseline system with a Bayesian weight update mechanism that individually calibrates the confidence of each rule based on observed extraction outcomes;
  \item the further optimisation of system parameters using a genetic algorithm, which tunes the fragment discard threshold and the minimum fragment count for frequency rule activation;
  \item an experimental comparison of three system configurations-baseline, Bayesian-enhanced, and GA-optimised-in terms of precision, coverage, and processing time.
\end{itemize}

Several challenges are also considered throughout the work, including the handling of dynamic JavaScript-rendered pages, anti-bot protection mechanisms, and the normalisation of price values across different currencies and formats.

The remainder of this work is organised as follows. Section~2 presents the state of the art and introduces the main concepts related to web scraping and price extraction. Section~3 describes the construction of the proposed browserless scraper, including the rule system and the fragmentation pipeline and presents the Bayesian and genetic algorithm extensions. Section~4 presents the experimental evaluation and comparison between the three implemented configurations. Finally, the last sections discuss the conclusions, limitations, and possible directions for future work.

\section{State of the art}

\subsection{Definitions and basic concepts}

To understand the principles underlying web data extraction systems, it is necessary to introduce the basic concepts on which the entire architecture of data parsing systems is built.

Web scraping is the practical implementation of automated web data extraction using wrappers or scrapers. Modern web scrapers often simulate browser behaviour and user interactions in order to retrieve data from websites~\cite{gavade2025}. Numerous approaches to web scraping exist, ranging from classical rule-based methods to modern AI/ML-based techniques. However, collecting raw HTML content alone is not sufficient: a scraping system must also determine the semantic meaning of the extracted data-for example, differentiating whether a numerical value represents a product price, weight, rating, or quantity. This task is addressed by a related, but distinct field known as \textit{Web Information Extraction}.

Web Information Extraction is the process of identifying and extracting target information items from web pages~\cite{liuwebmining}. Closely related to this concept is Web Data Extraction, which refers to the extraction of structured data records from web pages generated from underlying databases and presented using fixed templates~\cite{liuwebmining}. Such extraction enables the integration of information from multiple web sources for applications including comparative shopping, business intelligence, and meta-search systems~\cite{liuwebmining}.

Web information extraction systems typically operate on different types of data. Structured data has a predefined and clearly organised structure: it is stored in tables with fixed rows and columns, where each element corresponds to a specific type and format. For example, data retrieved via SQL queries is highly structured and stored in relational tables~\cite{liuwebmining}. Unstructured data, by contrast, has no predefined model or storage structure and is typically presented in free form-such as arbitrary text, for which no structured query language analogous to SQL exists~\cite{liuwebmining}. Semi-structured data occupies an intermediate position: it lacks a rigid tabular schema, yet contains special markers, tags, or elements that help organise and describe its content. HTML code, for instance, belongs to this category: it contains tags that define a certain page structure and divide it into elements-headings, paragraphs, links-while lacking the strict relational schema of a database. On the Internet, a user may encounter all three data types simultaneously: structured tables, semi-structured pages, unstructured texts, and multimedia files-images, audio, and video~\cite{liuwebmining}.

AJAX (Asynchronous JavaScript and XML) is a web technology that enables a page to exchange data with a server in the background without a full page reload. Instead of returning a complete new HTML document, the server sends only the data required to update a specific part of the page. This mechanism is widely used in modern e-commerce interfaces, where product prices, availability, and recommendations are loaded dynamically in response to user interactions. From a web scraping perspective, AJAX-rendered content is invisible to methods that retrieve only the initial HTML response, since the target data does not exist in the page source at the moment of the HTTP request but is injected into the DOM at a later stage by JavaScript.

Information Retrieval (IR) essentially means searching for a set of documents relevant to a user query, followed by ranking the results by degree of relevance. The most common query format is a list of keywords, also referred to as terms. IR differs fundamentally from database retrieval via SQL queries, since data in databases is highly structured, whereas information in texts is unstructured~\cite{liuwebmining,sarawagi2007}.

In general terms, the web scraping process consists of three sequential stages: target site analysis, crawling, and data organisation~\cite{kazmali2025}. The first stage is the discovery and access of the required pages through crawling-a process in which automated bots sequentially follow links from page to page, building a queue of URLs for subsequent processing~\cite{schneider2025}. Access to a website is obtained via an HTTP request. Formally, an HTTP interaction represents a request-response exchange described by a tuple $(U, V, P, B)$, where $U$ is a parameterised URL; $V$ is an HTTP verb (GET, POST, etc.); $P$ is a set of request parameters in the form of (name, value) pairs, with the origin of each parameter taking one of the following values: header, path, query, or post. The collection of all HTTP interactions recorded during a single system run forms an \textit{HTTP trace}, which is subsequently analysed to derive data extraction rules~\cite{fayzrakhmanov2018}.

Once the page content has been retrieved, a parser comes into play-a software component that accepts an HTML document as input and returns its structured representation: embedded objects, frames, links, forms, and other elements~\cite{kazmali2025}. A typical parsing process involves lexical analysis of the markup, construction of a parse tree, navigation through nodes using CSS selectors or XPath, and extraction of target values~\cite{schneider2025}. The intermediate structural object in this process is the DOM (Document Object Model)-a hierarchical tree representation of the HTML page, whose nodes correspond to markup tags and attributes. Data extraction algorithms operate directly on DOM subtrees, iterating and filtering nodes to obtain the required information~\cite{ferrara2010}. The page rendering procedure required to build the DOM involves three steps: parsing the web page and constructing the DOM tree, applying CSS rules, and executing JavaScript-the last being the most computationally expensive operation, which often substantially alters the DOM structure of the page~\cite{fayzrakhmanov2018}.

The central component of any data extraction system is a wrapper-a procedure implementing a family of algorithms that locate the required information within a source, extract it from its unstructured representation, and transform it into structured data, merging and unifying it for further processing~\cite{ferrara2010}. Any data extraction system must support both wrapper generation and wrapper execution. The wrapper lifecycle begins with generation-whether manual using regular expressions, inductive, or semi-automatic through visual interfaces-and continues with execution and subsequent maintenance as the structure of target pages changes~\cite{ferrara2010}. Based on the mode of interaction with a website, two main types of wrappers are distinguished~\cite{fayzrakhmanov2018}. A \textit{visual wrapper} simulates user actions-clicks, form filling, button presses-by firing DOM events within an embedded browser rendering engine, offering high versatility but at the cost of significant time and network resources for page rendering. An \textit{HTTP wrapper}, by contrast, interacts directly with the server via HTTP requests, bypassing page loading and rendering entirely, which substantially reduces execution time and the volume of transmitted data~\cite{fayzrakhmanov2018}.

At the final stage, the extracted data undergoes transformation: cleaning, conflict resolution, structural normalisation, and schema matching across multiple sources, after which it is packaged into the target format-a relational database, XML, CSV, or another storage system-and delivered to the managing system for analytical, commercial, or other applied purposes~\cite{ferrara2010}.

\subsection{General classification of approaches}

There are four main types of approaches to extracting data from web sources:

\begin{itemize}
  \item classical rule-based methods, in which the rules for locating and extracting data from text or HTML are defined manually;
  \item browser simulation methods, which imitate browser behaviour;
  \item browserless methods, where data is parsed by interacting directly with the website's server via HTTP requests without using a browser;
  \item AI/ML-based methods, where parsing is performed by trained models rather than by direct data reading.
\end{itemize}

In classical rule-based methods, their core concept is that a
programmer manually writes rules using a specialised pattern language or
regular expressions. Typical search criteria include word boundaries, HTML
tags, and table structures~\cite{ferrara2010}. Regular expressions are one
of the oldest yet still effective formalisms: they define matching criteria
for strings and patterns in unstructured text, and a wrapper built on them
dynamically generates data extraction rules~\cite{ferrara2010}. Writing such
expressions manually, however, is a difficult task that demands considerable
domain knowledge, which is precisely why automated regex-generation techniques
based on genetic algorithms have been proposed as an alternative to manual
authoring~\cite{aslanyurek2024regex}. Yet regardless of whether the rules are
written by hand or generated automatically, regular expressions as a class
share the same fundamental drawback: any change to the HTML structure of a
website breaks the extraction logic and requires the expressions to be
rewritten, which makes this approach difficult to scale and maintain in
practice~\cite{ferrara2010}. That said, regular expressions are not the only
technique within this category: wrapper induction, visual-clue based
extraction, and automatic extraction also belong here and will be examined
later in this section.

Browser simulation methods, their concept is built on the use of a visual wrapper that reproduces a sequence of user actions-such as filling in a form or clicking buttons-by firing DOM events within an integrated rendering engine~\cite{fayzrakhmanov2018}. This makes it possible to work with dynamically generated pages, and wrapper generation under this approach is considerably simplified compared to manual rule writing. At the same time, browser initialisation and rendering together require substantial computational resources. This drawback became the motivation to search for new parsing techniques, which led to the development of browserless methods.

Browserless methods are based on the HTTP wrapper-a program that interacts directly with the website's server via HTTP requests similar to those a browser would issue, and extracts the required data directly from the raw server response, without loading, rendering, or executing JavaScript~\cite{fayzrakhmanov2018}. In other words, the browser as an intermediary layer is eliminated from the process entirely. As a result, the resources spent on data retrieval are substantially reduced. However, this approach has a fundamental drawback from the perspective of wrapper creation and maintenance. Writing an HTTP wrapper is far from a trivial task: it requires studying the browser traffic, understanding the sequence of data exchange between client and server, identifying the precise flow of relevant parameters, and only then writing a parameterised wrapper. Furthermore, HTTP wrappers are generally less robust to website structural changes than visual wrappers.

Methods based on artificial intelligence represent an alternative approach to data parsing. Whereas classical rule-based, browser simulation, and browserless methods always rely to some extent on a predefined page structure or a fixed sequence of HTTP interactions, ML-based methods enable the system to learn to identify these patterns autonomously, based on training examples and statistical regularities. Accordingly, data extraction rules are derived automatically during training on labelled page examples~\cite{ferrara2010}. However, achieving acceptable accuracy requires a large volume of labelled web pages, and the training data must include both positive and negative examples.

Thus, the four approaches reviewed above constitute the foundational methods for web data extraction. Each has its own strengths and weaknesses, and none of them is universal. Each method implies its own type of solution, evaluated from the perspective of implementation simplicity, performance, robustness to website changes, and resource requirements~\cite{lloret2020,ferrara2010,fayzrakhmanov2018}.

To grasp the differences in practice more clearly, examining the tools linked to each category is beneficial. Table~\ref{tab:libperf} summarises four widely used libraries across the classical and browser-based categories, comparing them by memory usage, CPU usage, and working time~\cite{dikilitas2024}.

\begin{table}[H]
\centering
\caption{Performance Comparison of Web Scraping Libraries~\cite{dikilitas2024}}
\label{tab:libperf}
\begin{tabular}{lccc}
\toprule
Library & Memory Usage (MB) & CPU Usage & Working Time (sec) \\
\midrule
Scrapy        & 2{,}400 & 2.8 & 8.1 \\
BeautifulSoup & 8{,}500 & 3.7 & 8.6 \\
Jsoup         & 2{,}150 & 1.5 & 7.3 \\
HtmlUnit      & 2{,}600 & 4.1 & 9.7 \\
\bottomrule
\end{tabular}
\end{table}

The results reflect the primary balance between parsing simplicity and JavaScript handling capability. The technical reasons behind these numbers, as well as the practical implications for each approach, are discussed in the corresponding sections below.

\subsection{Classical methods}

Information extraction methods can be broadly divided into rule-based and statistical approaches. Rule-based methods are driven by hard predicates, whereas statistical methods make decisions based on a weighted sum of predicate activations. Rule-based systems are easier to interpret and develop, but they handle noise in unstructured data less effectively. For this reason, rule-based approaches are more commonly applied in closed domains where the data structure is stable and human involvement is feasible, while statistical methods prove more effective in open domains~\cite{ferrara2010}.

Classical rule-based data extraction methods include regular expressions, XPath, CSS selectors, DOM traversal, and wrapper-based extraction systems. These methods rely on predefined rules that describe how to locate the required elements within a text or HTML document. Their main advantages are ease of implementation, high precision under stable conditions, and good interpretability. However, they have a significant drawback-fragility in the face of changes to the HTML structure of pages, which complicates their maintenance~\cite{ferrara2010}.

The literature identifies four main approaches to building wrappers within the rule-based paradigm~\cite{lloret2020}. The first is \textit{manual writing}, in which the developer defines the extraction rules by hand using pattern languages or regular expressions. The second is \textit{wrapper induction}, where rules are automatically derived from labelled page examples. A concrete example of this approach is the RoadRunner system~\cite{roadrunner2001}, which fully automates wrapper construction for data-intensive websites-that is, sites containing large, regularly structured volumes of data, typical of online stores. Unlike classical wrapper induction, where rules are derived from examples labelled in advance by a human, RoadRunner requires neither labelled data nor any user interaction. During its operation, the system analyses two HTML pages from the same website simultaneously and constructs a wrapper based on their similarities and differences, while the data schema is inferred together with the wrapper itself rather than being specified in advance. Mismatches between two pages of the same template are used as a signal to identify meaningful structural elements, while matches form the basis for constructing a common grammar. The authors formulate the task as the induction of a regular grammar for the HTML code of pages sharing the same template, and confirm the feasibility of the approach through experiments on real data-intensive websites.

The third is the \textit{visual-clue based approach}, in which the system records user actions in the interface-such as clicks and selections-and automatically constructs a wrapper from them. The fourth is \textit{fully automatic extraction}, where the system independently identifies structural patterns in pages without any human involvement.

BeautifulSoup and Jsoup handle DOM parsing well but cannot deal with dynamic content at all ~\cite{pant2024bs}. The choice of tool therefore depends substantially on the project requirements: BeautifulSoup and Jsoup provide high efficiency for straightforward parsing tasks where the target data is available in the static HTML response, while tools capable of JavaScript execution are required for sites that rely on client-side rendering~\cite{abodayeh2023bs}.

Scrapy sits in the middle, as it does not construct a complete DOM tree but employs XPath and CSS selectors on raw HTML, and can be supplemented with JavaScript support through external tools. Its resource profile is more nuanced than it might initially appear. On the one hand, Scrapy is considerably more memory-efficient than BeautifulSoup. It uses roughly $3.5\times$ less memory (2{,}400~MB vs.\ 8{,}500~MB), which is explained by the fact that Scrapy does not load the entire parsed HTML tree into memory at once, whereas BeautifulSoup constructs and holds the full DOM object in RAM~\cite{dikilitas2024}. On the other hand, compared to Jsoup, the lightest library in the comparison, Scrapy consumes nearly $1.9\times$ more CPU (2.8 vs.\ 1.5) and takes around 11\% longer per task (8.1~sec vs.\ 7.3~sec). This is the cost of Scrapy's built-in concurrency engine and request pipeline~\cite{dikilitas2024}. A representative
practical application of this architecture is given by a system that extracted
product information from an e-commerce website using CSS selectors applied
directly to the server response, without constructing or traversing a full DOM
tree; the extracted data was subsequently used to build a
product-classification ontology~\cite{elasikri2020scrapy}.

An example of a classical XPath-based implementation for price extraction is described by Vu and Nguyen~\cite{vu2011}. The implemented system consists of two components: a front-end and a back-end.

The front-end component comprises three modules.

\begin{figure}[H]
  \centering
  \includegraphics[width=0.85\textwidth]{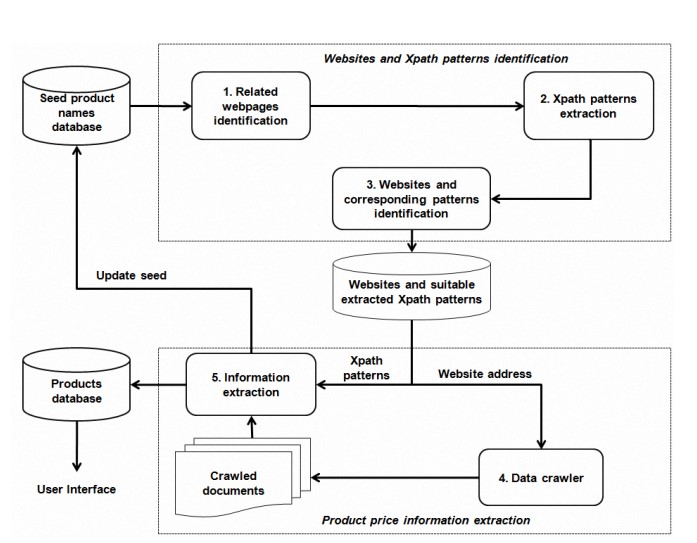}
  \caption{Architecture of price IR system~\cite{vu2011}}
  \label{fig:priceirsystem}
\end{figure}

\begin{enumerate}
  \item The relevant webpage identification module takes a set of seed product names as input and returns webpages associated with those names. Drawing on the specific characteristics of commercial websites, the system automatically constructs Google search queries corresponding to product names using predefined templates. For instance, instead of querying ``ipad~2'', the system automatically generates the query ``ipad~2'' $+$ ``VND or USD''.

  \item The XPath pattern extraction module receives a product name and its associated webpage from the previous module as input, and outputs the actual price along with XPath patterns for identifying product names and prices. Since pages within the same website typically share a common structure, these patterns can be reused to extract product names and prices from other pages of that site. The module itself consists of two sub-modules (see Figure~\ref{fig:priceirsystem}).

  \item The website and pattern identification module detects websites that match the extracted XPath patterns for subsequent generation of product names and actual prices. The module counts the number of pages per website that share identical XPath patterns identified by the previous module. If this count exceeds a predefined threshold, the website is classified as commercial and the corresponding patterns are considered suitable.
\end{enumerate}

The first sub-module identifies, within the Document Object Model (DOM) tree corresponding to the raw HTML of the page, the leaf node containing the text string that matches the input product name. The XPath pattern is then constructed by tracing the traversal path from the root node of the DOM tree down to the identified leaf node.

The second sub-module first locates the leaf node in the DOM tree that contains the actual price string, and then constructs the corresponding XPath pattern. The candidate price node is identified in two stages: first, basic regular expressions are applied to detect all numeric strings on the page; second, from among the detected strings, those that may represent actual prices are selected based on prefix rules, suffix rules, and exclusion rules. A prefix rule applies when a number is preceded by the word ``Price'' (\textit{Giá}) or ``VN\DJ{}'' (Vietnamese dong). A suffix rule applies when a number is followed by ``VN\DJ{}'', ``USD'', ``\DJ{}'' (dong), or ``\$''. Exclusion rules apply when a string is preceded by words such as ``Old price'' (\textit{Giá cũ}). The final determination of the actual price among multiple candidates is made by verifying the proximity relationship between the product name and the price: it is assumed that the name and price reside in two adjacent nodes of the DOM tree. For example, if the first sub-module yields the XPath pattern \texttt{HTML $\to$ BODY $\to$ TABLE $\to$ TR $\to$ TD $\to$ DIV[1] $\to$ product\_name}, and one of the candidate prices corresponds to the pattern \texttt{HTML $\to$ BODY $\to$ TABLE $\to$ TR $\to$ TD $\to$ DIV[2] $\to$ FONT $\to$ product\_price}, then the similarity measure equals 5 overlapping steps. The pattern with the highest similarity measure relative to the product name pattern is selected as the output pattern for actual price extraction.

The author reports results on a per-module basis. The XPath pattern extraction module demonstrated high accuracy across all three tested products. For Nokia~1200, the F-measure ranged from 93.88\% to 100\%; for Canon~G10, from 90\% to 100\%; while Lenovo~T61 yielded the lowest results, with an F-measure in the range of 89.89-90\%. The information extraction module produced uneven results depending on the structural properties of each website. On \texttt{www.dienthoaididong.com.vn}, 743 out of 792 commercial pages were successfully processed, yielding an accuracy of 93.81\%. On \texttt{www.trananh.vn}, the result was considerably lower-416 out of 711 pages (58.5\%).

These experimental results support the key claim reflected in Table~\ref{tab:libperf}: classical methods based on XPath and DOM parsing are capable of achieving high extraction accuracy-up to 100\% F-measure for Nokia~1200 and Canon~G10-while requiring minimal computational resources. The system ran on 2011-era hardware: an Intel Celeron 2.66~GHz processor with 768~MB of RAM~\cite{vu2011}.

At the same time, the accuracy drop to 58.5\% on \texttt{www.trananh.vn} clearly points to an inherent limitation of the classical approach~\cite{vu2011}: XPath patterns are rigidly tied to the structure of a specific website, and as soon as a site employs different HTML templates across product categories, the system begins to behave inconsistently.

It can therefore be concluded that classical methods are best suited to structurally uniform websites with predictable HTML markup and static content, making them an effective tool for price monitoring under resource-constrained conditions. However, scaling to heterogeneous sources requires either expanding the set of XPath patterns~\cite{vu2011}, or transitioning to the browserless or browser-based approaches.

\subsection{Methods that simulate browsing}

Browser simulation methods operate through a full browser or emulate user behaviour. Common tools in this category include Selenium, Puppeteer, Playwright, and headless browsers. Such tools are capable of rendering JavaScript, interacting with dynamic pages, and supporting user actions such as clicks, form filling, and page scrolling via DOM events (\textit{click}, \textit{keypress}, and others)~\cite{fayzrakhmanov2018,kazmali2025}. This makes it possible to work with content that is generated dynamically by JavaScript and AJAX after the page has loaded-content that classical extraction methods are unable to reach.

Browser automation tools can be divided into two subcategories. Full browser automation, as implemented by Selenium WebDriver, uses a real browser to access a website. Headless browsers, by contrast, support JavaScript and DOM operations in the same way as regular browsers but without a visual interface, which reduces resource consumption and speeds up the scraping process since there is no rendering step involved~\cite{kazmali2025}.

The key advantage of browser simulation methods is that they enable work with dynamic pages and significantly simplify wrapper creation. User actions can be recorded through a visual point-and-click interface, after which the system automatically generates extraction rules without requiring manual HTML analysis. This reduces development time from several days to a matter of hours~\cite{fayzrakhmanov2018}. Tools such as Mozenda, iMacros, Lixto, and Visual Web Ripper all operate on this principle~\cite{lloret2020,fayzrakhmanov2018}.

This effectiveness is grounded in a fundamental observation: while the internal implementation of a webpage-DOM node identifiers, CSS classes, JavaScript code-changes continuously, the user-facing interface tends to remain stable, as developers must preserve a consistent experience for their users. User-facing interface elements tend to remain stable across redesigns because developers must preserve a consistent experience, whereas the underlying DOM attributes change frequently.

However, this approach has a fundamental drawback that manifests at the execution stage. Page rendering in a browser involves three steps: constructing the DOM tree, applying CSS rules, and executing JavaScript, with the last step being the most resource-intensive~\cite{fayzrakhmanov2018}. Measurements show that browser initialisation accounts for approximately 13\% of total execution time, rendering for approximately 85\%, and only 2\% is attributed to the actual data extraction~\cite{fayzrakhmanov2018}. In other words, initialisation and rendering together require roughly 50 times more time than the data retrieval itself. Moreover, approximately 96.8\% of the internet traffic generated by a visual wrapper is entirely irrelevant from the perspective of obtaining the target data, as it consists of loading images, fonts, stylesheets, and advertisements~\cite{fayzrakhmanov2018}.

Thus, browser simulation methods solve the fragility problem of classical rules with respect to dynamic content, but introduce a new bottleneck in the form of rendering computational cost.

One example of a browser simulation method based on recording user actions is the Ringer system~\cite{ringer2016}. The authors proposed an approach to web automation that eliminates the need for reverse engineering of the target page. Instead of writing scripts manually, the user simply demonstrates the required interaction, and the system automatically generates a replayable script. Ringer is implemented as a Chrome browser extension written in JavaScript. Each script consists of a sequence of statements of the form ``wait for condition $C$, then perform action $a$ on element $e$'' and includes three key components:

\begin{enumerate}
  \item Actions. The system records user interactions at the level of DOM events such as \textit{mousedown}, \textit{mouseup}, \textit{click}, \textit{keydown}, and others, which allows precise reproduction of fine-grained interactions. This is what distinguishes Ringer from tools like CoScripter, where actions are limited to predefined high-level operations.

  \item Element identification. Instead of fixed XPath expressions or identifiers, Ringer applies the SIMILARITY algorithm. During replay, for each candidate node $n_c$ from the DOM tree, a score is calculated reflecting the number of matches with the features of the original node $n$. More than a hundred features are used, including CSS attributes, coordinates, dimensions, text, and the structure of the surrounding subtree. The node with the highest score is selected. As an example, during 60 seconds of interaction with a page, 1{,}499 ID modifications and 2{,}419 class modifications were recorded without a single page reload. The similarity algorithm is robust to such changes, whereas approaches like iMacros and ATA-QV lose the node entirely when its key attribute changes.

  \item Synchronisation triggers. For example, on a product page the content is greyed out after a user action and the price updates only once the server response arrives. Without synchronisation, the script reads the previous price. Instead of searching for visual cues, of which there are more than 600 per single interaction, Ringer monitors HTTP server responses. The ADD\_TRIGGERS algorithm replays the script several times, aligns server responses across traces using hostnames and URL paths, and conservatively assigns to each action only those triggers that precede it in all successful traces.
\end{enumerate}

\begin{figure}[H]
  \centering
  \includegraphics[width=0.85\textwidth]{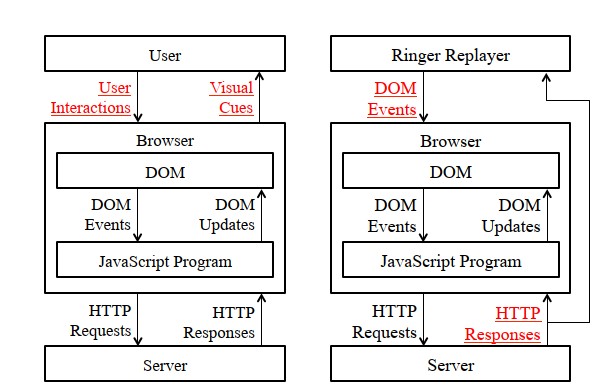}
  \caption{Ringer architecture~\cite{ringer2016}}
  \label{fig:ringer}
\end{figure}

The authors evaluated the system on a benchmark set of 34 interactions with real websites including Amazon, Gmail, Kayak, and Southwest Airlines from the Alexa ranking. Ringer successfully completed 25 out of 34 benchmarks (74\%), while CoScripter completed only 6 (18\%). Script robustness to page changes was tested over a three-week period, and out of 24 initially successful scripts, 20 produced at least one successful replay on every single test date, with only 2 experiencing persistent failures until the end of the period. In terms of replay speedup, the three-trace versions achieved an average speedup of $2.6\times$ compared to the user's original pace.

A related tool in this category is HtmlUnit, a Java-based library that simulates a browser environment without launching an actual browser window. Unlike Selenium or Playwright, which control real browsers, HtmlUnit implements its own lightweight engine capable of parsing HTML and executing JavaScript internally. This makes it faster than full browser automation while still handling dynamically generated content. However, as shown in Table~\ref{tab:libperf}, this comes at the cost of the highest CPU consumption among all four libraries tested (4.1), with a working time of 9.7 seconds per task-approximately 33\% slower than Jsoup and 20\% slower than Scrapy~\cite{dikilitas2024}.

\begin{table}[H]
\centering
\caption{Comparison of Browser Simulation Tools for Web Scraping}
\label{tab:browsersim}
\small
\begin{tabular}{p{1.6cm} p{3.2cm} p{3.0cm} p{1.4cm} p{3.2cm} p{1.4cm}}
\toprule
Tool & Language & Protocol & JS Exec. & Multi-browser & Speed \\
\midrule
Selenium   & Python, Java, C\#, Ruby & WebDriver (HTTP)      & Yes          & Chrome, Firefox, Safari, Edge & Slow     \\
Puppeteer  & JavaScript / Node.js    & Chrome DevTools (WS)  & Yes          & Chrome/Chromium only          & Fast     \\
Playwright & Python, JS, C\#, Java   & Chrome DevTools (WS)  & Yes          & Chrome, Firefox, WebKit       & Fastest  \\
HtmlUnit   & Java                    & Own HTTP client       & Yes (limited)& None (own engine)             & Moderate \\
\bottomrule
\end{tabular}
\end{table}

The tools in this category differ substantially in both architecture and performance. Selenium, the oldest of the four (in use since 2004), remains the standard choice in enterprise environments due to its support for multiple programming languages and all major browsers. However, it communicates with the browser via the WebDriver protocol, which introduces additional HTTP overhead, making it the slowest option in this group~\cite{apify2026}.

Puppeteer, developed by Google and released in 2017, replaced the WebDriver connection with a direct Chrome DevTools Protocol link over WebSocket, which significantly reduced communication overhead and improved execution speed. Its main limitation is that it supports only Chrome and Chromium-based browsers~\cite{apify2026}.

Playwright, released in 2020 by Microsoft, extended the DevTools approach to cover multiple browsers including Firefox and WebKit. It also introduced browser contexts-lightweight isolated sessions that can be spawned within a single browser instance without launching multiple processes-which makes it particularly efficient for large-scale parallel scraping. According to benchmarks, Playwright is the fastest of the three tools for scraping modern web applications~\cite{apify2026}.

In addition to the three tools discussed above, the authors also include Cypress in their comparison \cite{garcia2024browser}. Cypress differs structurally from other browser automation frameworks because, instead of communicating with the browser through an external protocol, it executes both the test code and the tested application within the same browser process. This architecture eliminates network communication overhead and improves execution speed, but it also limits certain operations, such as working with multiple browser tabs or switching between browsers during a test session.

The same comparison shows that the differences between these tools are not limited to performance alone. Both Cypress and Playwright provide automatic waiting mechanisms that ensure page elements are available before interaction, as well as built-in assertion capabilities for validating results. In Selenium and Puppeteer, such mechanisms often require additional manual configuration, which increases test complexity and may reduce the robustness of automated scripts.

An empirical comparison conducted across ten test scenarios on a real e-commerce application showed that none of the three tools dominates simultaneously in speed, CPU usage, and memory consumption~\cite{moncpanczyk2025}. Playwright consumed the least CPU in most scenarios (e.g., 14.71\% versus 19.71\% for Cypress and 24.42\% for Selenium), yet Selenium consistently used the least RAM (26.55\% versus 28.88\% for Playwright and 33.01\% for Cypress) and in several scenarios was even faster than the other two tools, despite its WebDriver-based architecture. Cypress performed worst on memory usage in every tested scenario.

These performance findings are reinforced by reliability data, in a 24-hour continuous test, Selenium achieved 100\% uptime with no failures, while Playwright recorded 99.72\% uptime with four downtimes; Selenium's higher failure rate (0.1208-0.1336 failures per second) was attributed not to instability but to a larger number of test cycles completed in the same time window~\cite{almabruk2025reliability}. Taken together, these results suggest that tool selection should be guided not by raw speed alone, but by which resource-CPU, memory, or long-term stability-is critical for a given scraping pipeline.

One of the key practical challenges faced by browser simulation methods is the active blocking of automated requests by websites. This problem is examined in detail by Moskalenko et al.~\cite{moskalenko2019}, who developed a web scraping system with built-in anti-blocking capabilities based on Selenium WebDriver. The system is implemented in Python using Selenium WebDriver and includes six key components:

\begin{enumerate}
  \item The graphical interface, built with the Tkinter library, which allows the user to define scraping parameters including the target URL, CSS data selectors, number of iterations, and pause interval between requests.
  \item The Selenium WebDriver core. The system uses a full browser to send requests, making the program's activity indistinguishable from that of a real user.
  \item The IP rotation module. This functionality is implemented via a proxy server that changes the scraper's IP address at each iteration using a list of free anonymous proxies.
  \item The user agent rotation module. At each iteration, the browser's User-Agent string is changed, preventing identification of the scraper by client signature.
  \item The random events module. Using the \texttt{ActionChains} class from the Selenium library, random actions are generated including mouse movement, keystrokes, and page scrolling.
  \item The evaluation module. It verifies the correct rotation of the IP address and User-Agent string through external services, records the execution time of each iteration, and logs any errors that occur.
\end{enumerate}

The system was tested on the website \texttt{yandex.ru/sprav}. Results showed that without the anti-blocking modules, blocking began as early as the fifth iteration, with an average iteration time of 5.479 seconds. With the modules activated, no blocking occurred throughout the entire testing period, although the average iteration time increased to 6.862 seconds.

A practical continuation of the problem of detecting automated browsers is provided by the FP-STALKER study~\cite{vastel2018fpstalker}, which focuses on browser fingerprinting-a technique for tracking users without relying on cookies, based on a unique combination of browser and system parameters such as the User-Agent, screen resolution, list of plugins, Canvas, and WebGL. The authors show that a browser's fingerprint inevitably changes over time: at least one change occurs within the first day in 45.2\% of cases, and the 90th percentile of the time to the first change is only 13 days. Nevertheless, the proposed hybrid algorithm, which combines static rules with a random-forest-based model, is able to link successive fingerprints of the same browser for an average of 54.48 days, and for more than 100 days in the case of 26\% of browsers. This implies that even with regular rotation of the User-Agent and IP address, as applied, for example, in the system of Moskalenko et al.~\cite{moskalenko2019}, a website can still recognise the same browser through more stable attributes such as Canvas or WebGL, which considerably complicates the task of bypassing blocks for browser simulation methods.

\subsection{Browserless methods}

The main idea of browserless methods is that the system sends HTTP requests directly to the server, receives raw HTML or API responses, and extracts the required data without using a rendering engine. This category includes HTTP clients (such as the Python \texttt{requests} library), API scraping, lightweight parsers, and browserless architectures~\cite{lloret2020,fayzrakhmanov2018}.

An HTTP wrapper-the central tool of this approach-interacts directly with a remote web server by sending HTTP requests similar to those a browser would issue, and retrieves the desired data directly from the raw server response without rendering~\cite{fayzrakhmanov2018}. For competitive price monitoring applications in e-commerce, where data from hundreds of websites must be collected in near real time, this makes browserless methods the preferred approach~\cite{lloret2020,rayarao2025}.

The advantages of browserless methods include high performance, low computational overhead, and strong scalability. The disadvantages are associated with the difficulty of handling JavaScript-heavy sites, where content is generated dynamically on the client side and therefore absent from the raw server response. Additionally, bypassing anti-bot systems-CAPTCHAs, IP blocking, and browser fingerprinting checks-poses a significant practical challenge~\cite{schneider2025}. Furthermore, writing HTTP wrappers by hand is an extremely complex task: even experienced developers typically require several days to develop and test a single new wrapper~\cite{fayzrakhmanov2018}.

This problem is solved in industry on a massive scale by the residential IP proxy (RESIP) business, as documented by the Resident Evil study~\cite{mi2019residentevil}. RESIP providers are commercial services that give clients access to the IP addresses of ordinary home users, rather than data-center addresses, specifically to evade server-side blocking. The authors analysed five of the largest such providers-Luminati, Geosurf, ProxyRack, Proxies Online, and IAPS Security. A random-forest classifier confirmed that 95.22\% of these addresses are indeed residential, yet only 2.20\% had ever appeared on public blacklists and just 0.06\% on lists of public proxies, making them effectively invisible to standard reputation-based blocking systems. Moreover, 90\% of these IP addresses serve as proxies for only about 870 seconds before being rotated, meaning the address changes faster than it can be blacklisted, which renders IP-based blocking largely ineffective. Despite providers' claims that hosts join voluntarily, the authors identified 237,029 IoT devices (routers, webcams, DVRs) and 4,141 infected hosts being used as residential proxies without their owners' consent-indicating that a substantial part of this evasion infrastructure relies on compromised devices rather than willing participants. This shows that evading blocks in practice is rarely a matter of swapping a single IP address, but instead relies on a constantly rotating pool of short-lived, hard-to-blacklist addresses.

One architectural way to address this blocking challenge is the application of multi-agent systems (MAS). A multi-agent system consists of several autonomous agents that act on their own initiative, react to their environment, and communicate with each other to solve a shared task~\cite{wooldridge2009}. This approach replaces a single monolithic script with a distributed group of specialized agents that coordinate with one another.

The classical architecture of a MAS is the three-layer Presentation~-~Business Logic~-~Data Access model, which includes~\cite{blumlapshin2026}:

\begin{enumerate}
  \item The collection layer~- crawler agents that interact directly with data sources (web pages, APIs, sensors).
  \item The processing and coordination layer~- filtering, aggregation, data cleaning, and load distribution among crawler agents.
  \item The storage and access layer~- databases, IPFS, API interfaces for delivering processed data.
\end{enumerate}

The system includes the following types of agents~\cite{blumlapshin2026}:

\begin{itemize}
  \item the coordinator agent distributes tasks among crawler agents;
  \item the crawler agent downloads pages;
  \item the parser agent extracts structured data from HTML;
  \item the data handler agent performs cleaning, validation, and storage of data;
  \item the anti-block agent solves captchas and reacts to JS challenges.
\end{itemize}

This architecture provides parallelism and scalability during data collection. Numerical estimates are also available for the specific problem of bypassing blocks: a study focused on the IP-address and User-Agent rotation module~\cite{moskalenko2019} showed that without bypass mechanisms the scraper was already blocked at the 5th iteration, whereas with the bypass modules applied no blocking occurred, although the average iteration time increased by approximately 25\%. This result confirms that bypassing blocks requires additional processing time, a trade-off that should be accounted for when designing a MAS architecture intended to retrieve complete information from each target site.

Beyond the architectures discussed above, the Bayesian approach can be introduced to make rule-based browserless methods more adaptable to website changes. This is a statistical method based on Bayes' theorem,
which allows the probability of an object belonging to a given class to be computed
using data accumulated through observation~\cite{mackay2003information}. Unlike
classical rules with a fixed weight, a Bayesian model can account for system
uncertainty and update its estimates as new information becomes available.

One example of such an application is the text classifier, which uses word
frequency to compute the probability that a page belongs to a target category.
In the task of phishing site detection~\cite{zhang2011antiphishing}, a page is
converted into a histogram of term occurrences, after which a naive Bayes
classifier estimates the probability that a given combination of words
corresponds to a ``phishing'' page. This approach moves away from manually
setting the classification threshold and instead relies on a statistical
estimate, where the threshold is chosen by scanning the observed values in the
training set and selecting the one that maximises the posterior probability of
correct classification, which is equivalent to minimising the number of
classification errors. According to the experimental results obtained on eight
datasets, the Bayesian threshold estimate outperformed the fixed threshold on
most metrics. For instance, for HSBC the CCR rose from 99.22\% to 99.70\%, while
the number of missed phishing pages dropped from 34/226 to 9/226~\cite{zhang2011antiphishing}.

An important aspect of building a browserless solution is choosing the right
regular expressions to locate the relevant fragments containing the price. A
syntax error in a regex causes mismatches, and whenever the site structure
changes the pattern has to be rewritten. This problem can be addressed by
automatically generating regular expressions. One suitable method is the
genetic algorithm, which is particularly useful when accuracy matters more
than the speed of parameter selection, since it can reach more accurate
solutions than Random Search and Grid Search, although at the cost of longer
computation time~\cite{shanthi2023ga}, for example when extracting relevant
images from web pages~\cite{aslanyurek2024regex}. During the algorithm's run,
a population of candidate regular expressions is created and evolves until an
expression is found that identifies the relevant data with the highest
accuracy. In the final accuracy of relevant image
extraction reached $\approx$98.49\%, and the method was also faster than
classical web scrapers, since it works at the level of textual HTML features
without downloading the images themselves~\cite{aslanyurek2024regex}.

At the same time, the genetic algorithm does not always require more time than
alternative methods.  In a study by Alibrahim and Ludwig~\cite{alibrahim2021ga}, a genetic algorithm, a Bayesian algorithm, and Grid Search were compared for tuning the hyperparameters of a neural network. The genetic algorithm outperformed both other methods not
only in accuracy, reaching 0.9059 against 0.8976 for Grid Search and 0.8959 for
the Bayesian algorithm, but also in the time required, around 12 hours against
16.5 and 23 hours, respectively ~\cite{alibrahim2021ga}. Thus, both the genetic and the Bayesian methods
prove to be suitable tools for adapting data extraction rules, which makes it
reasonable to combine their efforts when building a browserless solution.

One application of the Browserless Methods approach is described in the paper ``A browserless architecture for extracting web prices'' by Lloret-Gazo~\cite{lloret2020}, where the task is to extract prices from web pages on which several numeric values may simultaneously be present in different formats and contexts.

\begin{figure}[H]
  \centering
  \includegraphics[width=0.85\textwidth]{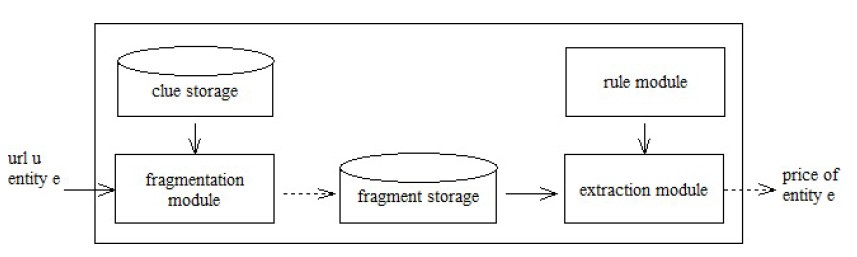}
  \caption{Browserless architecture for web price extraction~\cite{lloret2020}}
  \label{fig:lloretarch}
\end{figure}

The method consists of four steps. The first step is \textit{browserless extraction of raw HTML}, where the raw HTML code is retrieved directly from the page URL and passed to the architecture. The second step is \textit{fragmentation}, during which fragments of HTML code containing indicators of candidate prices are identified. The third step is \textit{rule application}, where a decision is made whether to discard or retain each candidate fragment. Rules are divided into three types: syntactic rules, which are based on the meaning of HTML elements (for example, the presence of a \texttt{<strike>} tag indicates a crossed-out price that is no longer valid); semantic rules, which are based on the meaning of words within a fragment, for example, the word ``Save'' indicates a discount rather than a price; and frequency rules, which are based on the repetition of patterns across multiple fragments.If the first $n$ characters are identical in three or more fragments, all of them are discarded. The fourth and final step is \textit{price extraction}.Intermediate data is stored in two relational tables — one for pages and one for candidate fragments. After the rules have been applied, a small number of fragments remain, and the target price is extracted from them. Figure~\ref{fig:lloretarch} illustrates all stages of the algorithm.

 An important aspect of the algorithm is the weight assignment technique applied to each candidate price fragment. Initially, every fragment receives a weight of~0. When a discarding rule is applied, the weight of a fragment increases by~1 if the condition is met, and remains unchanged otherwise. After all rules have been applied, fragments with a weight greater than zero are discarded, leaving only those whose weight remains~0. To prevent a situation where a rule eliminates all remaining candidates, each rule is equipped with a \textit{p\_limit} parameter: the rule is applied only if the number of still-available fragments is not less than this threshold. For most rules $p\_limit=0$, while for frequency rules it takes the value of~3.

The results of this technique are notable. The author states that the architecture demonstrates high performance across all 735 sites: average precision of 81\%, average specificity of 97.61\%, perfect results (100\% precision and specificity) for 577 out of 735 sites (78.5\%), and 100\% specificity for 630 out of 735 sites (85.71\%). The full experiment across 735 pages took 10 minutes, meaning the average price extraction time is 0.8 seconds per page.

\subsection{Methods based on artificial intelligence}

Methods based on artificial intelligence differ fundamentally from the three approaches above: instead of manually describing extraction rules, the system learns to find data from examples and statistical patterns~\cite{ferrara2010}. Whereas the previous methods rely on a predefined page structure or a fixed sequence of HTTP requests, AI/ML approaches change the fundamental operating principle: instead of manually describing extraction rules, the system learns to find data automatically based on examples and statistical patterns~\cite{ferrara2010}.

The conceptual foundation of these methods is the idea of wrapper induction, where extraction rules are not defined manually but are derived automatically from labelled training page examples~\cite{ferrara2010}. Contemporary AI/ML methods encompass several directions. Machine learning is used for pattern recognition and for analysing structural changes in pages~\cite{gavade2025}. NLP-based approaches are applied to text processing and to the extraction of named entities, semantic content, and sentiment from unstructured content~\cite{schneider2025}. Graph Neural Networks are used to analyse relational data~\cite{schneider2025}. In parallel, intelligent agents capable of autonomously adapting to website changes and bypassing anti-scraping mechanisms are being actively developed~\cite{schneider2025,kudratovich2025}.

\begin{figure}[h]
    \centering
    \includegraphics[width=0.9\textwidth]{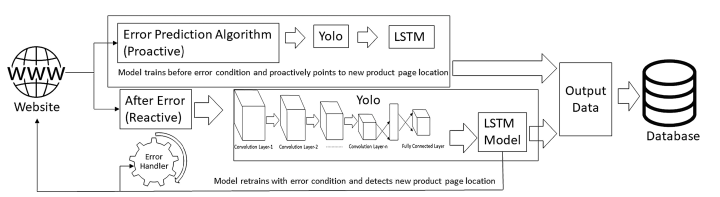}
    \caption{Architecture and offline experimentation environment of the
    proposed web information extraction framework~\cite{patnaik2022webextraction}.}
    \label{fig:ArchitectureLSTM25}
\end{figure}

Also, with an ML-based approach in the field of website parsing, it is possible
to address the problem of proactively predicting web page failures (layout
changes, URL migration, page unavailability) together with automated
information extraction, without the need to manually reconfigure the core
extraction engine~\cite{patnaik2022webextraction}. This system is built around
an LSTM network. LSTM (Long Short-Term Memory) is a type of recurrent neural
network (RNN) specifically designed to learn from long data sequences. Unlike
standard RNNs, LSTM uses a gating mechanism with input, forget, and output
gates, which regulate which information is kept in the cell's memory over the
long term and which is discarded~\cite{hochreiter1997lstm}. This network is
trained on historical website error data and predicts the error code for a
future period. When a potential failure is detected, the system constructs a
new predicted URL for the page. For data extraction, a combination of YOLO and
Tesseract/LSTM is used. YOLO is an architecture for real-time object detection
on an image which, unlike two-stage approaches, performs detection and
classification in a single pass over the network, considering the whole image
at once and predicting bounding boxes and object classes within a single
regression model~\cite{redmon2016yolo}. The reported results were 99\%
failure-prediction accuracy and 96.9\% overall information-extraction accuracy
even after the page structure changed. The system architecture is shown in
Figure~\ref{fig:ArchitectureLSTM25}. This example illustrates the general point
that AI/ML methods can automatically adapt to changes in page structure,
although this adaptability comes at the cost of significantly higher
computational and training-data requirements compared with rule-based
browserless approaches.

The primary advantage of AI/ML methods is their robustness to changes in web page structure, since they do not depend on hard-coded rules. However, these methods require large volumes of training data, substantial computational resources, and present high complexity in terms of both development and interpretability~\cite{ferrara2010}.

Within AI/ML methods, two practical strategies coexist. Wrapper-based
approaches still rely on fixed rule sequences and work well only on
sites with stable, predictable structures. Language-agent approaches
use LLMs to interpret free-form queries and extract data from both
structured and dynamic pages, but they depend on repeated API calls
to powerful models, which makes them slow and costly at scale.

Both approaches carry fundamental scalability limitations. Wrapper-based methods struggle considerably when confronted with unfamiliar site layouts. Language agents, despite their superior adaptability, impose significant time and financial costs due to their reliance on powerful API-based LLMs.

To address the shortcomings of both paradigms, the most effective solution is the paradigm of generating web scrapers using LLMs. The LLM is invoked only once to generate a scraper in the form of an XPath action sequence. The resulting scraper is standard executable code that can subsequently run across thousands of pages without any further calls to the LLM, making it substantially more resource-efficient.

The work in~\cite{autoscraper2024} describes the implementation of AutoScraper, a two-stage framework for progressive web scraper generation using LLMs. The framework produces a sequence of XPath expressions derived from a set of seed webpages. All expressions except the last are used for pruning the page, while the final expression extracts the target value from the pruned subtree. The model operates in two stages.

In the first stage, the algorithm generates XPath expressions by applying a top-down operation at each step, where the LLM progressively refines a path downward from the root node of the current DOM tree to the specific node containing the target information. The step-back operation is triggered when the XPath returns no result at all. In this case, the system moves up the DOM tree toward a more reliable and broadly applicable node to enable more precise XPath targeting. The maximum number of retry attempts is bounded by the parameter $d_{max}$.

The second stage is synthesis, since there is no guarantee that the XPath produced from a single page will generalise to other pages on the same site. During synthesis, $n_s$ seed webpages are selected at random and an individual action sequence is generated for each. The final XPath is then selected as the one that performs best across all pages in the sample.

\begin{table}[H]
\centering
\caption{Extraction performance (F1) on SWDE: supervised baselines vs.\ AutoScraper (zero-shot, GPT-4-Turbo)~\cite{autoscraper2024}}
\label{tab:autoscraper}
\begin{tabular}{lcc}
\toprule
Model & Type & F1 \\
\midrule
FreeDOM (Lin et al., 2020)       & supervised & 82.32 \\
Reflexion + GPT-4-Turbo          & zero-shot  & 82.40 \\
SimpDOM (Zhou et al., 2021)      & supervised & 83.06 \\
Render-Full (Hao et al., 2011)   & supervised & 84.30 \\
MarkupLM\textsubscript{BASE} (Li et al., 2022) & supervised & 84.31 \\
WebFormer (Wang et al., 2022)    & supervised & 86.58 \\
AutoScraper + GPT-4-Turbo & zero-shot & 88.69 \\
\bottomrule
\end{tabular}
\end{table}

AutoScraper was evaluated on the SWDE dataset covering 80 websites across 8 domains. With the best-performing backbone, GPT-4-Turbo, AutoScraper achieved a correct extraction rate of 71.56\%, compared to 61.88\% for COT and 67.50\% for Reflexion. The unexecutable rate dropped to 4.06\% versus 14.37\% for COT. The F1-score reached 88.69\%, against 76.95\% for COT and 82.40\% for Reflexion. Notably, AutoScraper operating in zero-shot mode, without any training data, outperformed all five supervised baselines. This highlights the framework's strong generalisation capacity and reduced resource requirements.

As the number of seed webpages increases from 1 to 5, the correct extraction rate rises consistently for both models: GPT-4-Turbo climbs from approximately 65\% to 73\%, while GPT-3.5-Turbo improves from 44\% to 57\%. At the same time, the unexecutable rate decreases steadily. However, the accuracy gains diminish with each additional seed page, suggesting that the performance benefits of expanding the seed set are subject to an upper bound.

Another practical example of the language-agent-based approach is the WebArena environment~\cite{zhou2024webarena}-a fully autonomous, reproducible web environment for building and testing LLM agents on applied tasks. Unlike AutoScraper, which generates an executable XPath script in a single pass, WebArena requires the agent to interact with the environment repeatedly in real time: the environment comprises four fully functional websites-e-commerce, a discussion forum, a collaborative software development platform, and a content management system. The results show that this task remains extremely difficult for current LLMs: the best GPT-4-based agent achieved an end-to-end task success rate of only 14.41\%, whereas humans handle the same tasks in 78.24\% of cases-indicating a substantial gap between agent and human performance. Of the 61 templates, GPT-4 achieved full success on only four, while the weaker GPT-3.5 model failed to fully solve any template. Similar results are shown by the Mind2Web benchmark~\cite{deng2023mind2web}, in attempts to scrape the site for new tasks within familiar websites (Cross-Task), new websites within familiar domains (Cross-Website), and entirely new domains (Cross-Domain). In the course of its work, the best model achieved a step success rate-the proportion of steps on which both the element and the action were correctly selected simultaneously-of 52.0\% in the Cross-Task scenario, but only 38.9\% and 39.6\% in the Cross-Website and Cross-Domain scenarios respectively, meaning accuracy drops noticeably when moving to unfamiliar websites and domains. Moreover, the proportion of tasks completed entirely without a single error across all steps was only 5.2\% even for the best model in the simplest scenario. This shows that continuous, real-time site navigation is fundamentally harder than the one-off extraction of structured data from a page, which is what AutoScraper specialises in.

One direction aimed at closing this gap between step-level accuracy and overall task success, revealed in the works on WebArena and Mind2Web, is the move toward multimodal agents. WebVoyager~\cite{he2024webvoyager} is a web agent built on a large multimodal model that interacts not with a simulator or a textual representation of the DOM, but directly with screenshots of real, live websites, on which interactive elements are marked with numerical labels overlaid on the image itself. The authors compiled a benchmark of real-world tasks from 15 popular websites and showed that WebVoyager achieves a 59.1\% task success rate, whereas a purely text-based GPT-4 (All Tools) agent achieves only 30.8\%, and a text-only variant of WebVoyager itself, lacking visual input, achieves 40.1\%. Thus, moving from a purely textual representation of the page to a combination of screenshot and text nearly doubles the proportion of successfully completed tasks. This indicates that the limitations of text-only agents observed earlier on Mind2Web and WebArena are explained not only by the inherent difficulty of multi-step navigation, but also by the lack of visual information about the page that humans naturally rely on when performing the same tasks.

The VIPS (Vision-based Page Segmentation) algorithm~\cite{vips2003} is described as an automatic approach to web page segmentation based on visual representation, operating independently of the underlying DOM tag tree. The core premise is that page designers naturally organise content so that semantically related elements are visually grouped together.

The algorithm proceeds through three sequential stages. In the first stage, visual blocks are extracted from the DOM tree using four heuristics: tag cues, colour cues, text cues, and size cues. In the second stage, visual separators are identified between the extracted blocks, and each separator is assigned a weight based on inter-block spacing, differences in font size and weight, and background colour contrast. In the third stage, blocks are merged into a hierarchical structure starting from the lowest-weight separators, continuing until the user-defined coherence threshold $PDoC$ is reached.

For each block, a Degree of Coherence metric (DoC) is defined, ranging from 0 to 1, reflecting the semantic homogeneity of the block's content. The algorithm was evaluated on 140 web pages drawn from 14 main categories of the Yahoo directory. Of these, 86 pages (61\%) were rated as perfectly segmented, 50 pages (36\%) as satisfactory, and only 4 pages (3\%) failed, yielding an overall accuracy of 97\%. VIPS demonstrates that relying on visual cues such as colour, font, size, and spatial layout enables semantically meaningful page structure to be recovered independently of HTML implementation~\cite{vips2003}. However, the algorithm does not identify what specifically within a block constitutes a price or other target value-it is better understood as a preprocessing step that drastically narrows the search space before extraction. The actual retrieval of specific values can then be carried out using XPath expressions, regular expressions, or NLP techniques~\cite{weerasinghe2024}.

\subsection{Comparative analysis of all instruments}

The analysis of data parsing algorithms identified four main categories: classical methods, browser simulation methods, browserless methods, and methods based on artificial intelligence. All of them address the data extraction problem, but each one has its own specific approach to parsing. To make the comparison between them more precise, Table~\ref{tab:comparison} groups the key indicators of each approach:
\begin{itemize}
    \item {peak accuracy} — the best result of the method;
    \item {accuracy under structural variability} — the result on heterogeneous or changing websites;
    \item {computational cost} — the time and resources required to process a page;
    \item {test sample size} — the volume of data on which the method was evaluated by its authors.
\end{itemize}

\begin{table}[H]
\centering
\caption{Comparison of the four price-extraction methods.}
\label{tab:comparison}
\small
\begin{tabular}{|p{2.2cm}|p{2.6cm}|p{2.4cm}|p{2.8cm}|p{2.4cm}|p{2.2cm}|}
\hline
{Method} & {Representative system} & {Peak accuracy} & {Accuracy under structural variability} & {Computational cost} & {Test sample size} \\
\hline
Classical & Vu \& Nguyen~\cite{vu2011} & F1 = 89.89--100\% & 58.5\% (drop of $\sim$41 p.p.) & Low (2011-era hardware) & 3 product categories \\
\hline
Browser simulation & Ringer~\cite{ringer2016} & 100\% working scripts initially & 83\% (20/24) after 3 weeks & High: 85\% of time spent on rendering; 96.8\% of traffic useless~\cite{fayzrakhmanov2018} & 34 real-world benchmarks \\
\hline
Browserless & Lloret-Gazo~\cite{lloret2020} & 78.5\% of sites with perfect results & Average precision 81\%, specificity 97.61\% & Low: 0.8 s / page & 735 heterogeneous sites \\
\hline
Artificial intelligence & AutoScraper (GPT-4-Turbo)~\cite{autoscraper2024} & F1 = 88.69\% zero-shot & 88.69\% retained across 8 domains without retraining & High: depends on the cost of LLM calls; unexecutable rate 4.06\% & 80 sites, 8 domains \\
\hline
\end{tabular}
\end{table}

{Robustness to structural variability of websites} was chosen as the unified comparison metric, that is, the ability of the method to preserve its accuracy when moving from a homogeneous to a heterogeneous sample of pages. This metric can be traced across all four approaches and is the most representative for practical e-commerce price monitoring tasks, where the scraper must work with hundreds of different marketplaces.

The XPath-based system of Vu and Nguyen~\cite{vu2011} is the main representative of classical methods, and it achieves the highest accuracy of all four approaches on homogeneous websites, with an F1 score ranging from 89.89\% to 100\% across three product categories. For example, when extracting prices from a single electronics store where all product cards share the same HTML template, the system returns a practically error-free result, and it does so on comparatively modest hardware - the original evaluation ran on 2011-era machines. This changes once the site's structure varies: applying the same rules to different HTML templates within the same site drops accuracy to 58.5\%, a loss of roughly 41 percentage points~\cite{vu2011}. In other words, a single structural change to the target site is enough to invalidate the extraction rules entirely.

Browser simulation methods, represented by the Ringer system~\cite{ringer2016}, were evaluated in two complementary experiments. In a longitudinal robustness test, out of 24 initially successful scripts, 20 remained functional throughout three weeks of testing as the target sites evolved ($\sim$83\%), with only 2 scripts experiencing persistent failures. Separately, on a broader one-off benchmark of 34 real-world sites, Ringer completed 25 (74\%), compared to only 6 (18\%) for the CoScripter baseline. When parsing sites with dynamically loaded content, classical methods fail, whereas Ringer correctly reproduces user actions. Among the advantages of this approach is its ability to handle dynamic content and AJAX interactions inaccessible to classical methods, as well as the possibility of bypassing some anti-bot defenses. The drop in accuracy and efficiency, in turn, occurs under high load: rendering accounts for around 85\% of total execution time, and 96.8\% of the generated network traffic carries no useful data~\cite{fayzrakhmanov2018}. In addition, the study by Moskalenko et al.~\cite{moskalenko2019} confirmed that each additional protection module increases the iteration time from 5.479 to 6.862 seconds, that is, by approximately 25\%.

Browserless methods, represented by the architecture of Lloret-Gazo~\cite{lloret2020}, eliminate rendering overhead entirely and achieved an average precision of 81\% and an average specificity of 97.61\% across 735 sites, with 78.5\% of sites yielding perfect results and an average processing time of 0.8 seconds per page. A characteristic application scenario is the massive collection of prices from hundreds of online stores simultaneously for a competitor monitoring system, where a browserless scraper processes pages dozens of times faster than a browser simulator. The strengths of the approach include low resource consumption, high speed, scalability, and accuracy comparable to that of classical methods, achieved on a significantly more heterogeneous sample of 735 sites, whereas Vu and Nguyen evaluated their system on only 3 product categories. This means that the typical performance of browserless on diverse data is comparable to the performance of the classical method in its most unfavorable conditions, which indicates substantially better robustness to structural variability. The accuracy drop, however, occurs on JavaScript-rendered sites, where the content is generated on the client side and is absent from the raw HTML response of the server; in addition, the development and maintenance of HTTP wrappers require considerable effort.

Methods based on artificial intelligence, represented by AutoScraper built on GPT-4-Turbo~\cite{autoscraper2024}, achieved an F1 score of 88.69\% in zero-shot mode across 80 sites from 8 domains, outperforming five supervised models trained on labelled data. The unexecutable rate - the proportion of cases in which the extraction code generated by the model fails to run due to syntactic or logical errors and therefore produces no result, it was only 4.06\%, compared to 14.37\% for the baseline COT method. A typical application scenario is launching a scraper on a new, previously unseen marketplace without any manual rule configuration, in which case the model itself determines where the price is located on the page and constructs the extraction logic. The main strength of the approach is its superior generalization: it adapts to structurally heterogeneous sources without manual rule definition and retains 88.69\% F1 across 8 different domains without retraining. The efficiency drop, however, occurs when working with small sites fewer than $\sim$20 pages, where the cost of generating the scraper is not offset by the volume of extracted data, and also under budget constraints, since each LLM call has a cost and the method becomes expensive at large scale. In addition, the quality of the result depends directly on the capabilities of the underlying language model.

A direct numerical comparison of the methods is complicated by the fact that the original works use different metrics and samples of different sizes. Nevertheless, when peak and ``variability'' accuracy are compared, a clear pattern emerges: the higher the peak accuracy, the more pronounced its drop on heterogeneous data. Classical methods are the extreme case of this pattern, with a fall from 100\% accuracy down to 58.5\%, while artificial intelligence methods are the opposite, with 88.69\% retained across 8 domains without retraining. Browserless methods occupy an intermediate position, combining acceptable accuracy (81\%) with minimal computational cost (0.8 s / page), which makes them the preferred choice for large-scale e-commerce price monitoring~\cite{lloret2020, rayarao2025}.

Looking at the overall picture, no single method is universally optimal, and the choice of approach is determined by the specific task. Classical methods lose when there are many structurally heterogeneous sites, since each change in markup requires rewriting the rules, and the approach therefore does not scale~\cite{vu2011}. Browser simulation loses on tasks involving a large number of pages under limited computational resources: 85\% of the time is spent on rendering, and parsing hundreds of sites in real time becomes infeasible~\cite{fayzrakhmanov2018}. Browserless loses on sites with dynamic JavaScript content, where the data is simply absent from the server's HTML response, and also where complex anti-bot defenses must be bypassed~\cite{lloret2020}. Artificial intelligence methods lose on small-scale tasks and under strict budget constraints, since the cost of LLM calls makes them inefficient for one-off or small data collections~\cite{autoscraper2024}.

Thus, classical and browserless approaches are well suited to stable, high-volume tasks under resource constraints; browser simulation is justified when the main obstacle is dynamic content or anti-bot protection; and artificial intelligence methods provide the best scalability across heterogeneous sources but require greater computational resources and depend on the capabilities of the language model. For the practical task of this work, namely monitoring prices on hundreds of heterogeneous marketplaces under limited computational resources, the optimal choice is the browserless approach, which justifies its use as the foundation of the system developed in the second part of this work.

\section{Build a scraper browserless}

In this section, we present our main contribution: an adaptive browserless price-extraction system, developed and tested on more than 250 different marketplaces. We designed the system around the browserless approach, in which the system interacts directly
with the web server through HTTP requests and analyses the returned HTML
without rendering the page in a browser~\cite{fayzrakhmanov2018}. This
strategy requires considerably fewer computational resources than methods
based on full browser simulation. Empirical measurements reported by
Fayzrakhmanov et al.~\cite{fayzrakhmanov2018} show that HTTP wrappers are,
on average, $23.8$ times faster than their visual counterparts, and that
approximately $96.8\%$ of the network traffic generated by a visual wrapper
is irrelevant for price extraction, since it corresponds to product images,
fonts, stylesheets and advertising banners that do not contribute to the
target data. For these reasons, the browserless approach was adopted as the
foundation of the price-scraping system developed in this work.

Nevertheless, the rule-based extraction logic employed in this browserless implementation relies on marketplace-specific rules, a limitation independent of whether a browser is used, since browser-based methods rely on similarly structured selectors. As a consequence, rules calibrated for one
marketplace often cease to perform effectively on others, and their
accuracy degrades whenever a website's layout is updated. To address this
limitation while preserving the computational advantages of the
browserless approach, the baseline price-extraction method is extended
with two adaptation mechanisms: Bayesian rule-weight updating and a
genetic algorithm. First, each extraction rule is assigned an individual
confidence coefficient, which is refined as the system accumulates
experience through the Bayesian approach. Then, two global parameters of
the system, which affect all rules simultaneously, are tuned automatically
using the genetic algorithm. The joint use of these two methods is
justified by the fact that both have proven to be effective tools for
adapting data-extraction rules and improving the robustness of scraping
to changes in website structure while preserving high extraction
accuracy~\cite{aslanyurek2024regex, alibrahim2021ga}.
\subsection{General structure}

\begin{figure}[H]
    \centering
    \includegraphics[width=0.95\textwidth]{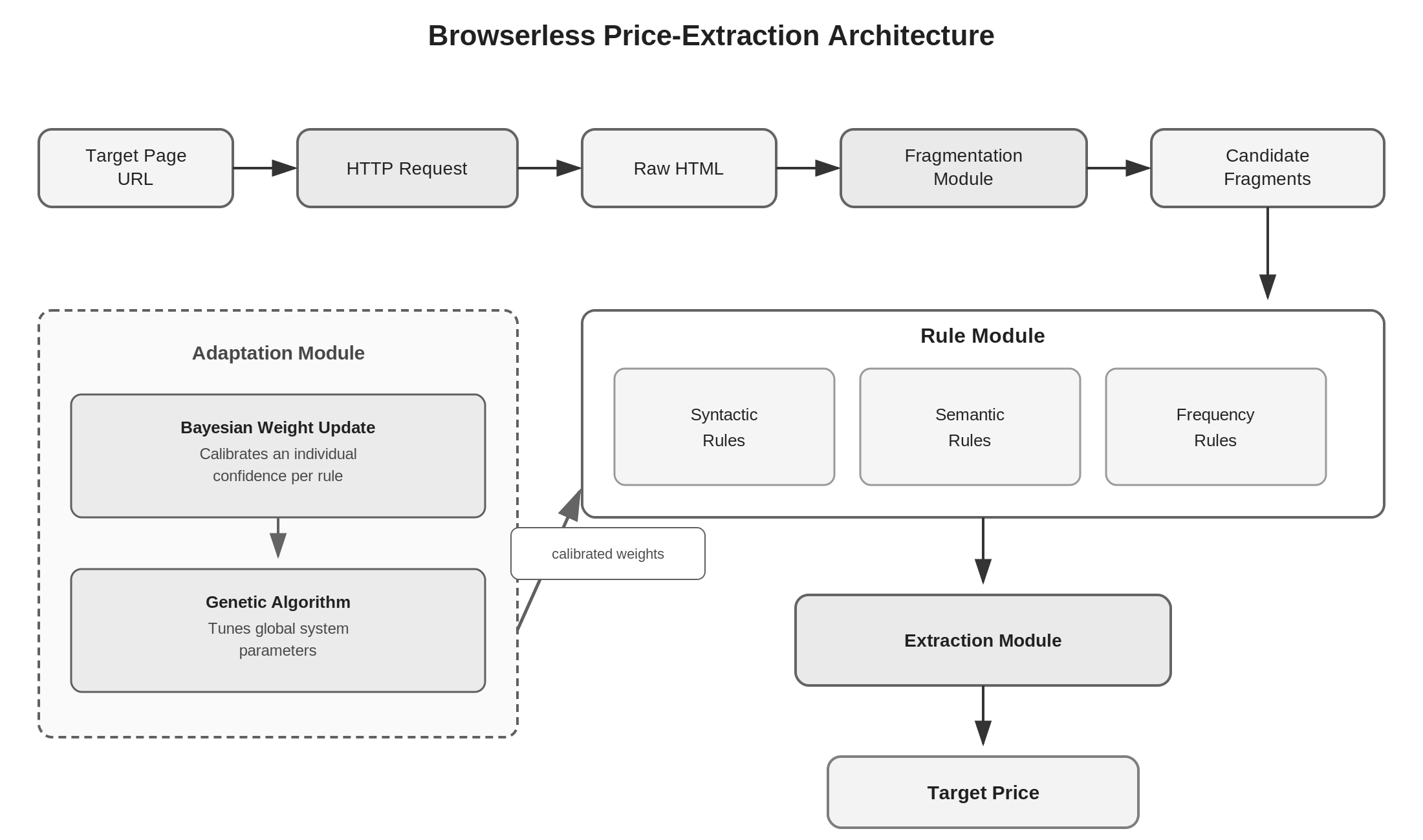}
    \caption{Architecture of the browserless price-extraction system: the price-extraction pipeline (HTTP request, fragmentation, rule application, extraction) and the adaptation module (Bayesian weight update and genetic algorithm).}
    \label{fig:architecture}
\end{figure}

This section describes the basic architecture of the browserless scraper,
on top of which the Bayesian weight update and the genetic-algorithm
parameter optimisation are subsequently built. The system consists of two
logical components: the price-extraction module, in which HTML retrieval,
fragmentation, and rule application take place, and the adaptation module,
combining Bayesian rule weight updating and genetic parameter optimisation.
At the final stage, all three variants of the system - the baseline, the
version with Bayesian updating, and the version with Bayesian updating and
the genetic algorithm - are compared with each other on the test set in
terms of precision, coverage, and processing speed. The overall
architecture is shown in Figure~\ref{fig:architecture}, and the stages of
the algorithm are as follows.

First, the system retrieves the HTML code of the page directly via an HTTP
request (see the HTTP Request block in Figure~\ref{fig:architecture}), without using a browser and without executing JavaScript. The
request is sent with headers that mimic a regular browser, in order to
reduce the risk of being blocked.

Next, elements that contain both a currency indicator and a number matching
the price format are extracted from the page's HTML code. Each candidate element is referred to as a \emph{fragment of level 1},
the HTML element whose text contains both a currency clue and a
numeric pattern compatible with a price. Its parent element is stored
separately as a \emph{fragment of level 2}, which
provides the contextual information, such as class and identifier attributes,
used at a later stage to disambiguate between several surviving candidates.

Three types of rules are applied to each fragment. Syntactic rules rely on
the HTML structure surrounding the fragment, for example the presence of
strikethrough tags or an input element, image, or script. Semantic rules
analyse the text adjacent to the detected number, responding to keywords
indicating, for instance, a discount or an old price before a discount.
Frequency rules penalise values that recur more often than others, which is
typical of advertising or navigation elements.

\begin{algorithm}[H]
\caption{Generic discarding rule}
\label{alg:rule}
\begin{algorithmic}[1]
\Procedure{Rule}{$p_{\text{condition}}$, $p_{\text{fragments}}$, $p_{\text{confidence}}$, $p_{\text{limit}}$}
    \If{$|p_{\text{fragments}}| \geq p_{\text{limit}}$}
        \For{each fragment $f$ in $p_{\text{fragments}}$}
            \If{$p_{\text{condition}}(f)$}
                \State $f.\text{weight} = f.\text{weight} + p_{\text{confidence}}$
            \EndIf
        \EndFor
    \EndIf
\EndProcedure
\end{algorithmic}
\end{algorithm}

A rule contributes to a fragment's accumulated weight only when its
condition is satisfied; if the condition does not hold, the fragment is
left unpenalised. The contribution itself is not a fixed constant but a
confidence value $p_{\text{confidence}} \in [0,1]$, learned and continuously
updated through the Bayesian mechanism described in Section~4.2. Before any
Bayesian training has taken place, this value is initialised at $0.5$,
reflecting complete uncertainty about the rule's reliability. The parameter
$p_{\text{limit}}$ ensures that a rule is only applied when a sufficient
number of fragments remain - which is particularly relevant for frequency
rules - and is initialised to a default value of $3$.

If the accumulated penalty exceeds the discard threshold, the fragment is
deemed irrelevant and excluded from further consideration. Fragments for
which no rule has fired retain a penalty of zero and remain the primary
candidates at the next stage; it is among these - denoted $F$ - that the final price is sought
first. For a fragment $f$, $\text{price}(f)$ denotes its associated price;
when the set of candidates contains a single fragment, its price is taken
directly as the result.

If exactly one fragment survives after the rules have been applied, its
price is taken as the final result. If several candidates remain, or none
at all, the system sequentially applies additional rules. First it looks
for a fragment with an explicit semantic marker of the current price -
formally, the predicate $\text{marker}(f)$, which holds when $f$ contains
such a marker - then for a fragment whose parent element's class name
indicates that it is specifically the product price, expressed by the
predicate $\text{class}(f)$. If none of these methods yields a single
result, the most frequently occurring value among the remaining candidates
is returned, denoted $\text{mode}(\cdot)$. 

\begin{algorithm}[H]
\caption{ExtractionModule}
\label{alg:extraction}
\begin{algorithmic}[1]
\Procedure{ExtractionModule}{$F$}
    \If{$|F| = 1$}
        \State \Return $\text{price}(F)$
    \EndIf

    \State $E_1 \gets \{f \in F : \text{marker}(f)\}$
        \Comment{marker $\sim$ \texttt{price}}
    \If{$|E_1| = 1$} \Comment{unique high-confidence candidate}
        \State \Return $\text{price}(E_1)$
    \EndIf

    \State $E_2 \gets \{f \in F : \text{class}(f)\}$
        \Comment{parent \texttt{class}/\texttt{id} $\sim$ \texttt{price}}
    \If{$|E_2| = 1$} \Comment{unique candidate by class/id heuristic}
        \State \Return $\text{price}(E_2)$
    \EndIf

    \If{$F \neq \emptyset$}
        \State \Return $\text{mode}(\text{price}(F))$
    \Else
        \State \Return \textsc{null}
    \EndIf
\EndProcedure
\end{algorithmic}
\end{algorithm}

The pipeline described above constitutes the baseline price-extraction
method and does not by itself include any training. Two adaptation
mechanisms are layered on top of it, operating in a separate cycle on the
training set. First, the Bayesian weight update is applied: for each rule, a
confidence coefficient is individually calibrated based on actual
experience - how often the firing of that rule led to a correct or an
incorrect result. Then a genetic algorithm is used, which, on top of the
already calibrated weights, automatically optimises two global parameters of
the system: the minimum number of fragments required to activate the
frequency rules, and the fragment discard threshold.

Once the rule weights have been calibrated by the Bayesian method and the
parameters tuned by the genetic algorithm, the resulting configuration is
applied to the extraction pipeline and evaluated on the test set. At this
stage, the three variants of the system - the baseline method without
adaptation, the method with Bayesian updating, and the method with Bayesian
updating combined with genetic optimisation - are compared, in order to
assess the accuracy gain provided by each added adaptation mechanism. The
following sections examine in more detail the Bayesian rule-weight-updating
algorithm and the genetic parameter-optimisation algorithm.

\subsection{Implementation tools}

The system was implemented in Python 3. This programming language is a de facto standard for web-scraping and data-processing tasks due to the large number of open-source libraries available and the relative ease of understanding its code~\cite{moskalenko2019, mitchell2015}. In comparative studies of scraping tools, the Python stack based on BeautifulSoup and Scrapy consistently appears among the most widely used solutions, both in academic and industrial data-extraction tasks~\cite{dikilitas2024, pant2024bs, abodayeh2023bs}.

The requests library is used as the HTTP client, providing HTML retrieval without JavaScript rendering, which is consistent with the browserless paradigm and aligns with the advantage of classical approaches in terms of memory and CPU consumption relative to browser-based solutions, as shown in Section 2.3~\cite{dikilitas2024}. Parsing of the retrieved HTML and the search for fragments around price clues are implemented using BeautifulSoup, a library specifically designed for extracting data from HTML documents through CSS selectors and DOM-tree navigation~\cite{moskalenko2019, abodayeh2023bs}. Pandas is used for handling the labelled dataset; the genetic algorithm and the Bayesian module are implemented using standard language features, without external ML frameworks, which preserves the transparency and reproducibility of the experiments.

\subsection{Bayesian weight update}

In the classical method, each rule contributes a fixed weight of one to the
accumulated fragment penalty~\cite{mackay2003information}. This means that
all rules are treated as equally reliable regardless of their actual
behaviour on a specific dataset. The present system extends this approach
with a Bayesian updating mechanism: each rule is assigned an individual
confidence coefficient, which adapts during training based on accumulated
experience.

The principle behind the rule weights is that the fragment weight represents
an accumulated penalty rather than a quality score. An initial value of zero
means that no rule has fired and the fragment is a candidate for price
extraction. Each fired rule adds its confidence value to the weight.

If the accumulated penalty exceeds the discard threshold, the fragment is
considered non-target and excluded from further processing. Thus, rules with
a high confidence value exert a strong influence on discarding, whereas
rules that frequently produce false positives gradually lose their
influence.

Each rule is initialised with identical counters: the number of correct
discards and the number of false discards are both set to one, which
corresponds to a confidence of 0.5 - complete uncertainty. Initialising the
counters at one rather than zero is a form of Laplace smoothing, which
prevents the confidence from collapsing to the extreme values of 0 or 1 with
few observations. Rule confidence is computed as the mean of the posterior
Beta distribution:

\begin{equation}
\text{confidence} = \frac{\text{true\_discards}}{\text{true\_discards} + \text{false\_discards}}
\end{equation}

\begin{itemize}
    \item {confidence} - the resulting confidence coefficient for the
    rule, a number between 0 and 1. The higher the value, the more reliable
    the rule: its firings, on average, more often lead to correct price
    extraction.
    \item {true\_discards} - the counter of ``correct discards'': the
    number of cases in which the rule penalised a fragment that did not
    contain the target price, and the final price was nonetheless correctly
    extracted.
    \item {false\_discards} - the counter of ``false discards'': the
    number of cases in which the rule penalised a fragment that actually
    contained the correct price, causing the system either to fail to extract
    the price or to extract an incorrect value.
\end{itemize}

After each page from the training set has been processed, the system
compares the extracted price with the ground-truth value and updates the
counters according to the following rules:

A \emph{correct discard} means that the rule penalised a fragment that does
not contain the correct price, and the system ultimately extracted the
correct price. The true\_discards counter is incremented by one, and
confidence increases; in subsequent iterations the rule will apply a
stronger penalty.

A \emph{false discard} means that the rule penalised a fragment that
contained the correct price, as a result of which the system failed to
extract it or extracted an incorrect value. The false\_discards counter is
incremented by one, and confidence decreases; in subsequent iterations the
rule will apply a weaker penalty.

Consider the rule responding to
the keyword ``shipping''. At the start of training its confidence is 0.5. As
the training set is processed, the counters evolve as shown in
Table~\ref{tab:confidence-evolution}.

\begin{table}[H]
\centering
\caption{Evolution of confidence for a semantic rule responding to ``shipping''}
\label{tab:confidence-evolution}
\begin{tabular}{lp{4.5cm}cccc}
\toprule
Iteration & Situation & true & false & confidence & Outcome \\
\midrule
Start & Laplace prior & 1 & 1 & 0.500 & Neutral \\
1 & Correctly discarded ``Free shipping \$5.99'' & 2 & 1 & 0.667 & Strengthened \\
2 & Correctly discarded ``Shipping: \$8.00'' & 3 & 1 & 0.750 & Strengthened \\
3 & Falsely discarded ``\$95.79 + free shipping'' (price was in this fragment) & 3 & 2 & 0.600 & Weakened \\
4 & Correctly discarded ``Standard shipping \$4.99'' & 4 & 2 & 0.667 & Stabilised \\
\bottomrule
\end{tabular}
\end{table}

The rule stabilises at a confidence of approximately 0.65-0.70, which
reflects its actual reliability, since the word ``shipping'' almost always
indicates a delivery cost, although in rare cases the product price and the
delivery information appear within the same HTML fragment.

The interaction between confidence and the discard threshold determines
which combination of rules causes a fragment to be discarded. For a discard
threshold of 0.5, Table~\ref{tab:threshold-effect} illustrates several
cases.

\begin{table}[H]
\centering
\caption{Effect of confidence on the discard decision}
\label{tab:threshold-effect}
\begin{tabular}{lccc}
\toprule
Fired rules & Accumulated weight & Threshold & Decision \\
\midrule
shipping (conf=0.85) & 0.85 & 0.50 & Discarded \\
kg (conf=0.35) & 0.35 & 0.50 & Survived \\
from (conf=0.30) + off (conf=0.30) & 0.60 & 0.50 & Discarded \\
No rules fired & 0.00 & 0.50 & Survived \\
\bottomrule
\end{tabular}
\end{table}

These examples show that weak rules cannot discard a fragment on their own -
several rules with low confidence must fire simultaneously. This property
protects the system against isolated false positives.

The system includes three groups of rules. The
first type, syntactic rules, analyse the HTML structure of the fragment
(Table~\ref{tab:syntactic-rules}).

\begin{table}[H]
\centering
\caption{Syntactic rules}
\label{tab:syntactic-rules}
\begin{tabular}{lll}
\toprule
Rule & Pattern & Interpretation \\
\midrule
syntactic\_0 & \texttt{<strike>} & Strikethrough text - old price \\
syntactic\_1 & \texttt{<s>} & Strikethrough text - old price \\
syntactic\_2 & \texttt{<del>} & Deleted text - old price \\
syntactic\_3 & \texttt{<s ...>} & Tag s with attributes - old price \\
syntactic\_4 & \texttt{<s\textbackslash n>} & Tag s with a line break \\
syntactic\_5 & line-through & CSS strikethrough - old price \\
syntactic\_6 & \texttt{<option>} & Dropdown list element - not the product price \\
syntactic\_7 & \texttt{<input>} & Input field - not the product price \\
syntactic\_8 & \texttt{<img>} & Image with a number - not a price \\
syntactic\_9 & \texttt{<script>} & Price inside a script - not in the DOM \\
syntactic\_10 & also-like & Related-products block - another product's price \\
\bottomrule
\end{tabular}
\end{table}

The second type, semantic rules, analyse the textual content of the fragment
(Table~\ref{tab:semantic-rules}).

\begin{longtable}{lp{5cm}p{5cm}}
\caption{Semantic rules}
\label{tab:semantic-rules}\\
\toprule
Rule & Keyword / pattern & Interpretation \\
\midrule
\endfirsthead
\multicolumn{3}{c}{\tablename\ \thetable\ -- continued from previous page}\\
\toprule
Rule & Keyword / pattern & Interpretation \\
\midrule
\endhead
\midrule
\multicolumn{3}{r}{Continued on next page}\\
\endfoot
\bottomrule
\endlastfoot
semantic\_0-2 & save, saving, savings & Discount - shows the savings amount, not the price \\
semantic\_3 & was & Old price before a discount \\
semantic\_4-5 & regular, msrp & Manufacturer's suggested retail price \\
semantic\_6-7 & rrp, list price & List price - not the actual price \\
semantic\_8-10 & listprice, list-price, originalprice & Variants of the list price \\
semantic\_11-12 & original price, old price & Price before a discount \\
semantic\_13-14 & before, compare at & Comparative price \\
semantic\_15-16 & discount, off & Discount indicator \\
semantic\_17-18 & bonus, reward & Bonus points or reward \\
semantic\_19-21 & fee, tax, vat & Additional charges - not the product price \\
semantic\_22-25 & delivery, shipping, dispatch, despatch & Delivery cost \\
semantic\_26 & handling & Order-handling cost \\
semantic\_27 & order & Total order cost \\
semantic\_28-30 & subscription, per month, per year & Recurring payment - not a one-time price \\
semantic\_31 & annual & Annual subscription cost \\
semantic\_32-35 & minimum, from, starting, as low as & Minimum or starting price of a range \\
semantic\_36 & each & Price per unit in a set \\
semantic\_37-38 & qty, quantity & Quantity - not a price \\
semantic\_39-41 & weight, kg, lbs & Product weight - not a price \\
semantic\_42 & gram & Weight in grams - not a price \\
semantic\_43-47 & rating, reviews, stars, points, credits & Rating or points - not a monetary value \\
semantic\_48-49 & gift card, voucher & Gift card or voucher \\
semantic\_50-51 & coupon, promo & Promo code or promotional offer \\
\end{longtable}

The third type, frequency rules, are activated only when at least a given
minimum number of active fragments are present on the page. The first such
rule penalises fragments whose price matches the most frequently occurring
value: a recurring price is very likely to be a navigation element (a price
in the header, the cart, or recommendations) rather than the main product
price. The second rule penalises fragments whose price matches prices found
in the related-products block.

The trained confidence values are passed to the next stage of the
experiment, where the genetic algorithm optimises the threshold parameters
of the system on top of the personalised rule weights. Thus, the Bayesian
update and the genetic optimisation work together: the former adapts the
strength of each rule, while the latter tunes the global parameters of the
system.

\subsection{Parameter optimisation with a genetic algorithm}

Although the Bayesian update sets the strength of each individual rule,
there remain two global parameters of the system - the minimum number of
fragments required to activate the frequency rules, and the fragment
discard threshold - which stay fixed and affect all rules simultaneously.
Searching for the optimal values of these parameters manually is
time-consuming, since their optimal values depend on the structure of the
specific dataset and on the current state of the Bayesian weights. A genetic
algorithm is used to tune them automatically~\cite{goldberg1989ga}.

The optimisation problem is formulated as the search for the pair of
parameters that maximises the precision of price extraction on the training
set:

\begin{equation}
\begin{split}
(p\_limit\_freq^{*},\ discard\_threshold^{*}) = {}\\
\arg\max_{p\_limit\_freq,\ discard\_threshold}\ \text{precision}(p\_limit\_freq,\ discard\_threshold)
\end{split}
\end{equation}

where \textit{p\_limit\_freq} is the minimum number of fragments required to
activate the frequency rules, and \textit{discard\_threshold} is the
accumulated-penalty threshold above which a fragment is discarded; precision
is computed as the proportion of correctly extracted prices among all pages
for which the system returned any numeric value, for a given pair of
parameters.

These parameters directly affect the balance between recall and precision.
Too low a value of the activation threshold for the frequency rules
increases the risk of erroneously discarding a unique price, whereas too
high a discard threshold makes the system more tolerant of noise, leaving
more candidates but increasing the chance of an incorrect extraction. The
genetic algorithm searches for a balance between these extremes.

Each individual - one of the candidate solutions that the algorithm
iteratively explores and improves - is encoded as a pair of values: the
minimum number of fragments for activating the frequency rules, and the
accumulated-penalty threshold. The chromosome is initialised randomly within
the allowed ranges:

\begin{equation}
p\_limit\_freq \sim \text{Uniform}(2, 10) \quad \text{(integer)}
\end{equation}
\begin{equation}
discard\_threshold \sim \text{Uniform}(0.3, \ 1.5) \quad \text{(real-valued)}
\end{equation}

where $\text{Uniform}(a, b)$ denotes a uniform distribution on the interval
from $a$ to $b$ - that is, any value within this range has an equal
probability of being chosen during initialisation. The fitness value - a
numerical measure of how well a given individual solves the task - is
stored directly with each individual and updated after every evaluation;
this value is the extraction precision, i.e. the proportion of correctly
extracted prices among all pages for which the system returned any numeric
value at all. The parameters of the genetic
algorithm used in the experiment are given in Table~\ref{tab:ga-params}.

\begin{table}[H]
\centering
\caption{Genetic algorithm parameters}
\label{tab:ga-params}
\begin{tabular}{lp{6cm}c}
\toprule
Parameter & Meaning & Value \\
\midrule
Population size & Number of individuals coexisting in each generation & 10 \\
Number of generations & How many times the population is successively updated during evolution & 8 \\
Crossover probability & Fraction of cases in which crossover is applied to a pair of parents, rather than simple copying & 0.7 \\
Mutation probability & Fraction of cases in which a random change is applied to an individual parameter & 0.2 \\
Elitism & Number of best individuals carried over unchanged to the next generation & 2 \\
Training-subset fraction & Fraction of the training set used for fitness evaluation & 20\% \\
\bottomrule
\end{tabular}
\end{table}

 A population of a given number of random
chromosomes is created. For each individual, the full price-extraction
pipeline is run on a subset of the data using the Bayesian weights trained
at the previous stage of the experiment, and the fitness value is computed.
The population is sorted in descending order of fitness.

At each generation, the two best individuals are carried over to the next
generation unchanged (elitism). The remaining slots are filled with
offspring produced by the following operators.

The method used to select parent individuals is tournament selection: three
individuals are drawn at random from the population, and the one with the
highest fitness is chosen as a parent. The operator is applied twice to
obtain two parents. Tournament selection is preferable to alternative
selection methods because it is robust to the scaling of fitness values and
does not require their additional normalisation~\cite{goldberg1989ga}.

New individuals are created through crossover - the exchange of parameter
values between two parents, occurring with a given probability. Since an
individual is described by only two values, crossover amounts to a full
exchange of these values: the first offspring receives one parameter from
the first parent and the second parameter from the second parent, while the
second offspring receives the opposite combination.

In addition, mutation is applied to each parameter separately with a given
probability - a small random change in the value, intended to broaden the
space of explored solutions: a random perturbation is added to the current
value, after which the result is rounded if necessary and clipped to the
allowed range.

Algorithm~\ref{alg:ga} formalises the evolutionary loop described above. At
each generation, every chromosome is evaluated on the training sample (line
4), the population is ranked by fitness (line 6), and the search terminates
early once perfect precision is reached (lines 7-9). The next generation
is built by carrying the $e$ best individuals over unchanged (line 10) and
filling the remaining slots with offspring produced by tournament
selection, crossover, and mutation (lines 11-15).

\begin{algorithm}[H]
\caption{GeneticOptimisation}
\label{alg:ga}
\begin{algorithmic}[1]
\Procedure{GeneticOptimisation}{$N$, $G$, $e$}
    \Comment{$N$: population size; $G$: generations; $e$: elitism}
    \State $P \gets$ $N$ random chromosomes \Comment{Eq.~(3)-(4)}
    \For{$g = 1$ {to} $G$}
        \For{each $c \in P$}
            \State $\text{fitness}(c) \gets \text{precision}(c)$
        \EndFor
        \State order $P$ such that $\text{fitness}(P_1) \geq \text{fitness}(P_2) \geq \dots$
        \If{$\text{fitness}(P_1) \geq 1.0$}
            \State {break}
        \EndIf
        \State $P' \gets \{P_1, \dots, P_e\}$ \Comment{elitism}
        \While{$|P'| < N$}
            \State $c_1, c_2 \gets$ \Call{TournamentSelect}{$P$}, \Call{TournamentSelect}{$P$}
            \State $(c_1', c_2') \gets$ \Call{Crossover}{$c_1, c_2$} {or} $(c_1, c_2)$
            \State $P' \gets P' \cup \{\Call{Mutate}{c_1'}, \Call{Mutate}{c_2'}\}$
        \EndWhile
        \State $P \gets P'$
    \EndFor
    \State \Return $\arg\max_{c \in P} \text{fitness}(c)$
\EndProcedure
\end{algorithmic}
\end{algorithm}

During the evolutionary
process, the algorithm performs a fixed number of generations or stops early
once a precision of 100\% is reached. The monotonic increase in the best
fitness is guaranteed by the elitism mechanism, whereby the best chromosome
never disappears from the population. The average fitness also increases,
indicating that the quality of the population becomes more uniform and that
diversity decreases towards the later generations.

As noted earlier, the genetic algorithm is run after the Bayesian training
has been completed, and it receives as input the already calibrated
confidence values for all rules. This means that the genetic algorithm
optimises the global parameters on top of the personalised weights. This
two-level structure provides a separation of responsibilities: the Bayesian
module adapts the strength of each rule to the specific dataset, while the
genetic algorithm tunes the overall aggressiveness of the system.

Once the genetic algorithm has finished, the best chromosome is used to
construct the final rule module, combining the Bayesian weights with the
optimised values of both parameters. This final module is evaluated on the
test set and compared with the baseline method and with the method using
only Bayesian updating without genetic optimisation.

\subsection{Experimental results}

Each of the three systems - the baseline, the version with Bayesian
updating, and the version with Bayesian updating and the genetic algorithm -
is evaluated on the test set using three measures:

\begin{equation}
\text{Precision} = \frac{\text{number of correctly extracted prices}}{\text{number of pages for which something was extracted}}
\end{equation}

Precision reflects how well the system distinguishes the correct price from
unrelated numbers on the page.

\begin{equation}
\text{Coverage} = \frac{\text{number of pages for which something was extracted}}{\text{total number of pages}}
\end{equation}

Coverage reflects the proportion of pages for which the system was able to
find any numeric value at all, regardless of whether it is correct.

\begin{equation}
\text{Avg time} = \frac{\text{sum of processing times for all pages}}{\text{total number of pages}}
\end{equation}

Average time is simply the mean processing time per page.

Before comparison, the extracted price and the ground-truth value are
brought to a common format: currency designations are removed, thousands and
decimal separators are unified, and the values are then compared as numbers
rounded to two decimal places. The complete dataset collected during this
work contains approximately 1{,}000 product links from e-commerce
sites. For controlled evaluation of extraction quality and for
assessing scalability and behaviour under load, subsets of up to
200~URLs were used. The final metrics of the three systems on the test
set are shown in Table~\ref{tab:final-results} for the baseline, the
Bayesian optimisation of extraction thresholds, and the combined
Genetic Algorithm with Bayesian weighting scheme.

\begin{table}[H]
\centering
\caption{Final results on the test set.}
\label{tab:final-results}
\begin{tabular}{lccc}
\toprule
System & Precision & Coverage & Avg.\ time (s) \\
\midrule
Baseline                                & 77.2\% & 98.75\% & 0.620 \\
Bayesian                                & 83.5\% & 98.75\% & 0.546 \\
Genetic Algorithm + Bayesian weighting  & 87.3\% & 98.75\% & 0.533 \\
\bottomrule
\end{tabular}
\end{table}

All three systems achieved a coverage of 98.75\% across all
configurations, since coverage is determined by the fragmentation
architecture itself rather than by threshold tuning.

The baseline method, without any additional training, achieved a
precision of 77.2\%, correctly identifying the price for 61 out of 79
extracted pages. The average processing time per page was
0.620~seconds. This result serves as the reference point for
evaluating the effectiveness of the proposed improvements.

Bayesian training raised the precision to 83.5\%. The average
processing time dropped to 0.546~seconds, which is explained by the
more confident discarding of non-target fragments by rules with
increased confidence: the system reaches a single candidate faster and
spends less time on the final price-selection stage.

The system combining Bayesian updating with the Genetic Algorithm
showed the same coverage as the Bayesian-only system, while raising
precision to 87.3\% and further reducing the average processing time
to 0.533~seconds. Thus, the average processing time per page is
reduced by approximately 14\%, while the precision gain in the
transition from Baseline to Bayesian is around 8.2\% relative to baseline, and from
Baseline to Genetic Algorithm with Bayesian weighting around 13.1\%,
relative to the baseline.

The parameters found by the Genetic Algorithm differed substantially
from the default values: the minimum number of fragments required to
activate the frequency rules increased from 3 to 9, while the discard
threshold decreased from 0.500 to 0.404, as shown in
Table~\ref{tab:ga-parameters}.

\begin{table}[H]
\centering
\caption{Parameters found by the Genetic Algorithm.}
\label{tab:ga-parameters}
\begin{tabular}{lccc}
\toprule
System & p\_limit\_freq & discard\_threshold & Fitness (train) \\
\midrule
Baseline                                & 3 (default) & 0.500 (default) & -    \\
Bayesian                                & 3 (default) & 0.500 (default) & -    \\
Genetic Algorithm + Bayesian weighting  & 9           & 0.404           & 0.721 \\
\bottomrule
\end{tabular}
\end{table}

The increase in the threshold number of fragments means that the
frequency rules now fire only when a sufficiently large number of
candidates is present on the page, which prevents the erroneous
discarding of a price on pages with few price-related elements. The
decrease in the discard threshold, in turn, makes the system stricter:
a fragment is discarded with a relatively small accumulated penalty,
which reduces the number of candidates that reach the final
price-selection stage and thereby decreases the overall processing
time.

For a web-scraping system, in addition to the accuracy of the system,
it is also necessary to evaluate its resource consumption and
execution time. A classical centralised scraper is a single sequential
script, whereas a distributed or parallel approach replaces it with a
pool of executors, each of which independently processes its share of
URLs and coordinates with the others toward the common goal of
efficient and reliable data
collection~\cite{blumlapshin2026}. The use of parallelism contributes
to scalability and fault tolerance of the system~\cite{wooldridge2009}.

To evaluate to what extent these theoretical benefits are realised in
our implementation, an experiment was conducted with three execution
modes:

\begin{itemize}
  \item {sequential} - URLs are processed one after another
  in a single execution thread;
  \item {parallel-2} - parallel execution with a pool of
  2 workers. A \emph{worker} here denotes an independent execution
  unit that takes a URL from the
  queue, downloads the page, runs the extraction pipeline on it and
  returns the result, working in parallel with the other workers;
  \item {parallel-3} - parallel execution with a pool of
  3 workers.
\end{itemize}

\begin{figure}[h]
\centering
\includegraphics[width=0.8\textwidth]{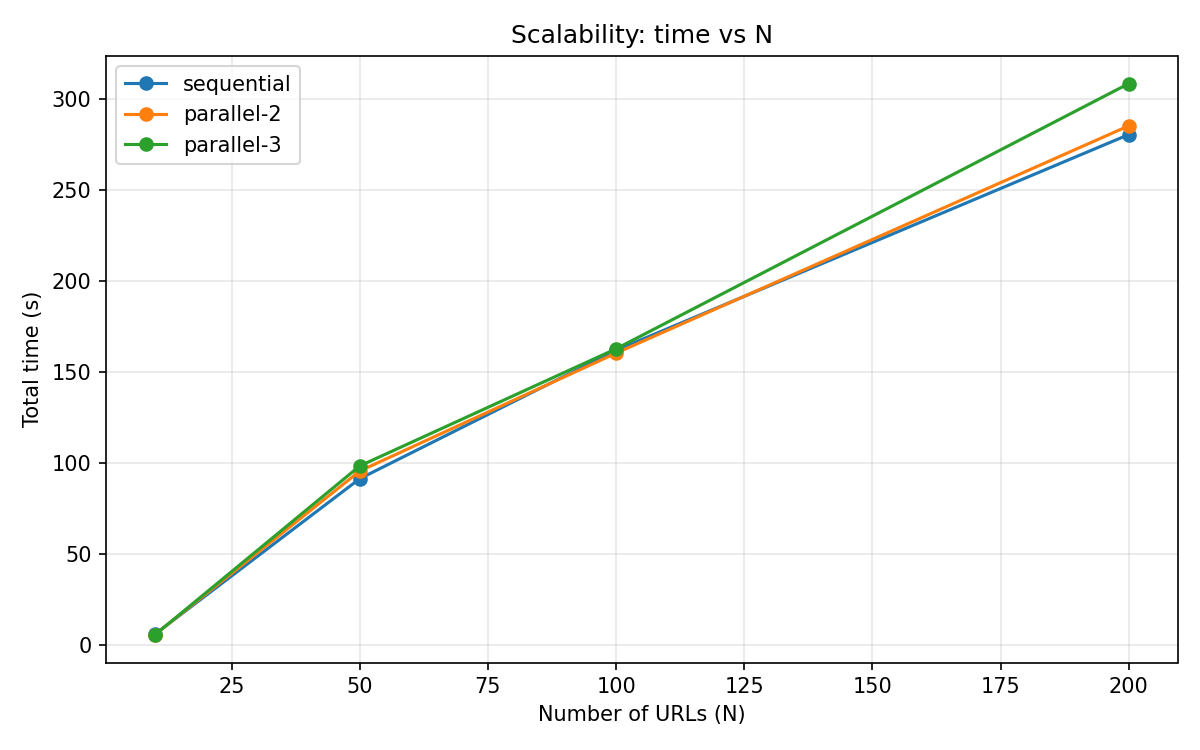}
\caption{Total processing time as a function of the number of URLs.}
\label{fig:scalability-time}
\end{figure}

\begin{figure}[h]
\centering
\includegraphics[width=0.8\textwidth]{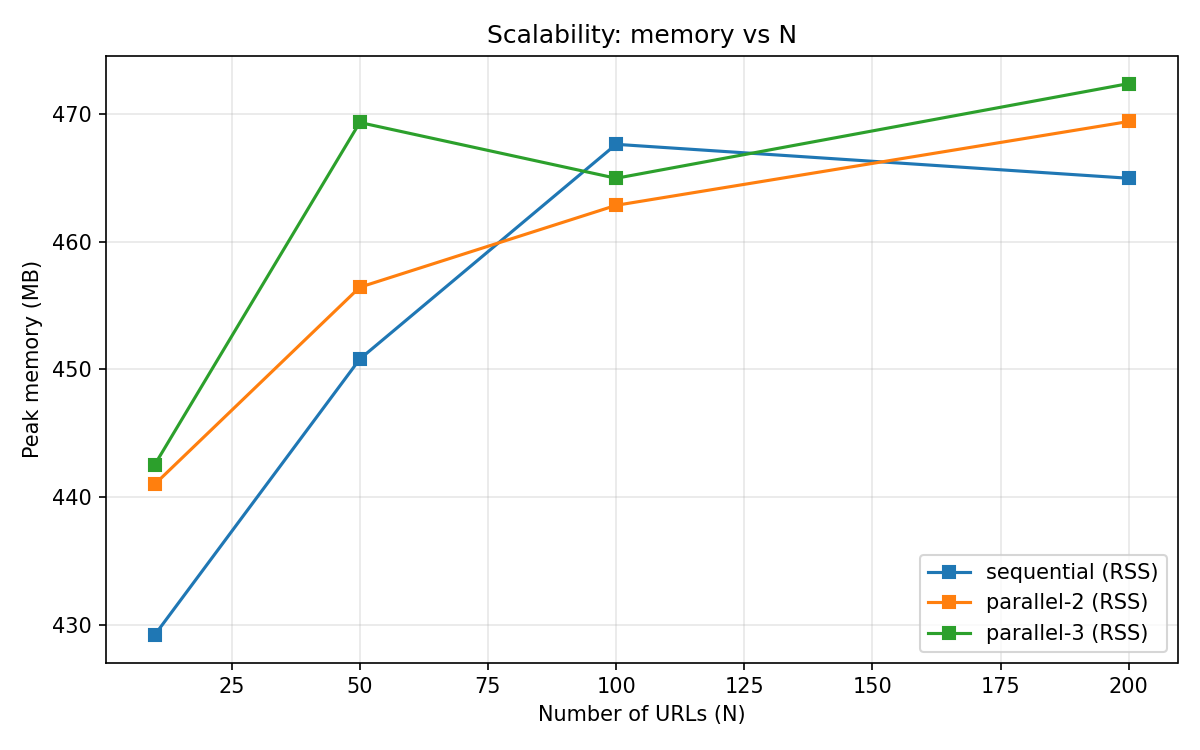}
\caption{Peak memory consumption (RSS) as a function of the number of
URLs.}
\label{fig:scalability-memory}
\end{figure}

Each mode was run on samples of size $N \in \{10, 50, 100, 200\}$
URLs. The total processing time
(Fig.~\ref{fig:scalability-time}) and the peak memory consumption
(RSS, Fig.~\ref{fig:scalability-memory}) were measured.

All three curves in Fig.~\ref{fig:scalability-time} are practically
linear in $N$ and lie close to one another. At $N = 200$, sequential
takes approximately 279~s, parallel-2 about 284~s and parallel-3
about 307~s. The closeness of the curves means that processing time
in the current task is determined primarily by the CPU stage of
extraction - HTML fragmentation, regular-expression generation and
rule evaluation - rather than by network waits, which usually
benefit from asynchronous execution. The linear growth pattern is
preserved across all three modes, which provides a predictable
processing time as the input volume increases.

Fig.~\ref{fig:scalability-memory} shows that peak
consumption depends only weakly on the number of URLs and is almost
independent of the execution mode. This means that memory is
determined mainly by the size of the model and the rule set, as well
as by HTTP-client buffers, rather than by the number of pages
processed simultaneously. The system does not accumulate the whole
batch in memory but processes pages in a streaming fashion. The
differences between modes are minimal (parallel-3 peaks at 472~MB
versus 465~MB for sequential at $N = 200$) and are explained by
per-worker buffers, which does not affect the overall memory
stability.

The experiments support two conclusions:
first, the hybrid scheme combining the Genetic Algorithm with the
Bayesian weighting system outperforms both the manually tuned
baseline and the purely Bayesian optimisation,
reaching a precision of 87.3\% at 98.75\% coverage while simultaneously reducing the average
per-page time - and this constitutes the main qualitative result of
the work. Second, the behaviour of the system under load is stable:
processing time scales linearly with the number of URLs, and peak
memory consumption stays within the narrow range of 430-470~MB
regardless of sample size or execution mode. Such linearity in time
and stability in memory is a suitable property for batch
data-collection systems, since it makes it possible to estimate in
advance the resources required to process an arbitrary number of
links. From a practical perspective, this means that the current
implementation, without any further modifications, is suitable for
processing the set of 1{,}000 links in a predictable time -
approximately 25~minutes by extrapolation from $N = 200$.


\subsection{Future work}

The results obtained confirm the reliability and sufficient accuracy
of the system. The main direction of further work is the
{extension of the experimental base}. In the present work, a
set of approximately 1{,}000 records was used, and with an 80/20
split the test sample amounts to about 200 records. In the future, it
is planned to double its size: after the downloading of the remaining
pages is completed, the test sample is expected to grow to
approximately 2{,}000 records, which will make it possible to obtain
statistically significant precision estimates with narrower confidence
intervals.

With an increase in the volume of data, the potential of the
individual modules of the system will also be revealed more fully.
The Bayesian module will receive an additional number of training
examples for differentiating the reliability of rules across
different product categories, while the Genetic Algorithm will be
optimised on a more heterogeneous sample, which is highly likely to
lead to parameters yielding a measurable accuracy gain relative to
the baseline configuration. Already at the current stage, the
Genetic Algorithm demonstrates a practical effect through the
reduction of the average per-page processing time - by
approximately 2.4\% relative to the Bayesian configuration (from
0.546~s to 0.533~s) - which indicates real changes in the system's
behaviour, in addition to the precision gain from 83.5\% to 87.3\%.

Beyond the extension of the dataset, the following directions of
development can be highlighted. For example, the {annotation
of the dataset by product categories and subsequent per-category
evaluation}. The current evaluation of the system is aggregated over
the entire test set and does not distinguish the specifics of
different types of products. Dividing the dataset into categories
such as electronics, clothing, books and food products, and
independently computing precision, coverage and average time for each
of them, will make it possible to obtain a more informative picture
of the strengths and weaknesses of the system. Categories with a
homogeneous page structure, for example books, will almost certainly
yield higher metrics than categories with diverse layouts and
non-standard price formats, such as household appliances with
discounts and instalment plans, and it is precisely these problematic
segments that should determine the priorities of subsequent
improvements.

Also, regarding the adaptation module of the system: in the current
implementation, the Bayesian rule weights are common for the entire
dataset, which averages the reliability of a rule across different
types of pages. After the per-category annotation it becomes possible
to maintain {separate sets of weights for each category}.
Category-dependent weights should increase extraction accuracy on
heterogeneous content and at the same time make the system's
behaviour more interpretable: for each category it will be possible
to identify which rules contribute most to the decision.

If one continues with the topic of the Bayesian weighting, an
{online update of the Bayesian module weights} could be
implemented in the future. In the current implementation, the
Bayesian weights are computed at the training stage and remain fixed
afterwards during exploitation. An autonomous system, however, may
continuously receive new pages, and the distribution of their
structures changes over time. The solution thus becomes a transition
to an online update scheme, in which each newly processed page
(together with a confirmed label about the correctness of the
extracted price) incrementally adjusts the posterior distributions
over the rules, without the need for full retraining. This will
allow the system to adapt to changes in websites in real time and to
maintain stable quality without periodic manual intervention.

Furthermore, one can consider a {comparison with alternative
baseline methods}. For instance, a comparison with fundamentally
different approaches: classical frameworks with static selectors,
such as Scrapy with manually written XPath/CSS rules; wrapper
induction methods, such as RoadRunner; and LLM-based parsers, which
use large language models to extract structured data from HTML. Such
a comparison will show under which conditions the proposed hybrid
scheme provides an advantage, and under which conditions alternative
methods remain a more suitable choice.

\section{Conclusions}

This work has presented, as its theoretical part, an in-depth review of the four
main approaches to web price extraction - classical, browser-based,
browserless, and AI/ML-based - and, as its practical part, a browserless
price-extraction system that combines a rule-based fragmentation pipeline with
two learning components: a Bayesian update of rule weights and a genetic
algorithm for tuning the extraction thresholds.

In the theoretical part of this work, a wide range of parsing implementations
was reviewed, from classical methods based on manually written rules to
solutions employing machine learning and large language models. The
comparative analysis showed that no single ideal solution exists: each
approach has its own advantages and limitations and proves optimal only under
a specific combination of conditions - data volume, available computational
resources, content dynamism, and the frequency of changes in website
structure. As a result, the choice of a price-extraction method should be
driven by the specific requirements of the task rather than by the universal
superiority of one approach over the others; it is precisely this conclusion
that justified the choice of the browserless approach as the foundation for
the practical part of this work.

In the practical part of this work, the three configurations were evaluated
comparatively on the same test set, which made it possible to determine the
contribution of each component. The main result was an increase in accuracy
when using the two adaptation mechanisms. Bayesian weighting alone increased
extraction precision by approximately 8.2\% relative to the manually tuned
baseline configuration, while the combined Genetic Algorithm and Bayesian
weighting scheme provided an increase in precision of approximately 13.1\%
relative to the baseline configuration. At the same time, the average
per-page processing time was reduced by approximately 14\% by the hybrid
scheme compared to the baseline. The behaviour of the system under load
proved to be stable: processing time scales linearly with the number of
URLs, and peak memory consumption remains within a narrow range regardless
of the sample size or execution mode, which makes the implementation
suitable for predictable batch processing.

As an overall outcome, these results confirm that the proposed
hybrid approach is an alternative both to manually tuned browserless
extractors and to more resource-intensive methods based on browsers
or large language models, offering competitive accuracy at low
resource cost. Nevertheless, the current results, obtained on a test
sample of approximately 200 records, should be regarded as a
reliable preliminary validation. As a continuation of the work, the
size of the test sample should be doubled, per-category evaluation
and category-dependent Bayesian weights should be introduced, a
transition to an online update of the weights should be made in
order to adapt to website changes in real time, and the proposed
scheme should be compared with alternative parsing methods, such as
frameworks with static selectors, wrapper induction systems and
LLM-based parsers, for a more precise determination of the operating
regime in which the hybrid approach offers the greatest advantage.
For reproducibility, the implementation, experimental scripts and
dataset are publicly available at
\url{https://github.com/Evgesha2022/TFM_web_parsing/tree/main}.

\newpage

\end{document}